\documentclass[fleqn,usenatbib]{mnras}

\usepackage{newtxtext,newtxmath}

\usepackage[T1]{fontenc}

\DeclareRobustCommand{\VAN}[3]{#2}
\let\VANthebibliography\thebibliography
\def\thebibliography{\DeclareRobustCommand{\VAN}[3]{##3}\VANthebibliography}

\usepackage{graphicx}	
\usepackage{amsmath}	

\title[Detectability of satellites around directly imaged exoplanets and brown dwarfs]{Detectability of satellites around directly imaged exoplanets and brown dwarfs}

\author[C. Lazzoni et al.]{
Cecilia Lazzoni$^{1,2},$\thanks{E-mail: c.lazzoni@exeter.ac.uk}
Silvano Desidera$^{2},$
Raffaele Gratton$^{2},$
Alice Zurlo$^{3,4,5},$
Dino Mesa$^{2}$
and Shrishmoy Ray$^{1}$
\\
$^{1}$University of Exeter, Astrophysics Group, Physics Building, Stocker Road, Exeter, EX4 4QL, UK\\
$^{2}$INAF -- Osservatorio Astronomico di Padova, Vicolo dell'Osservatorio 5, I-35122, Padova, Italy\\
$^{3}$N\'ucleo de Astronom\'ia, Facultad de Ingenier\'ia y Ciencias, Universidad Diego Portales, Av. Ejercito 441, Santiago, Chile\\
$^{4}$Escuela de Ingenier\'ia Industrial, Facultad de Ingenier\'ia y Ciencias, Universidad Diego Portales, Av. Ejercito 441, Santiago, Chile\\
$^{5}$Aix Marseille Univ, CNRS, LAM, Laboratoire d'Astrophysique de Marseille, Marseille, France
}

\date{Accepted XXX. Received YYY; in original form ZZZ}

\pubyear{2021}

\begin{document}
\label{firstpage}
\pagerange{\pageref{firstpage}--\pageref{lastpage}}
\maketitle

\begin{abstract}
Satellites around substellar companions are a heterogeneous class of objects with a variety of different formation histories. Focusing on potentially detectable satellites around exoplanets and brown dwarfs, we might expect to find objects belonging to two main populations: {\it planet-like} satellites similar to Titan or the Galileian Satellites - likely formed within the scope of core accretion; and {\it binary-like} objects, formed within different scenarios, such as disk instability. The properties of these potential satellites would be very different from each other. Additionally, we expect that their characterization would provide insightful information about the history of the system. This is particularly important for planets/brown dwarfs discovered via direct imaging (DI) with ambiguous origins. In this paper, we review different techniques, applied to DI planets/brown dwarfs, that can be used to discover such satellites. This was achieved by simulating a population of satellites around the exoplanet $\beta$~Pic b, which served as a test case. For each simulated satellite, the amplitude of DI, radial velocity, transit and astrometric signals, with respect to the planet, were retrieved and compared with the detection limits of current and future instruments. Furthermore, we compiled a list of 38 substellar companions discovered via DI to give a preliminary estimate on the probability of finding satellites extracted from the two populations mentioned above, with different techniques. This simplified approach shows that detection of {\it planet-like} satellites, though not strictly impossible, is very improbable. On the other hand, detection of {\it binary-like} satellites is within the capabilities of current instrumentation. 
\end{abstract}

\begin{keywords}
 planets and satellites: detection-- techniques:high angular resolution, image processing, photometric, radial velocities
\end{keywords}



\section{Introduction}

In the past decade, the direct imaging (DI) technique has proven to be an invaluable method to investigate planetary systems. In fact, even if the indirect methods such as radial velocities (RV) and transits are more prolific in terms of the number of detections, substellar companions detected via direct imaging have the advantage of revealing simultaneously a plethora of information on the object. The combination of 8-meters size telescopes with the performances of extreme adaptive optics devices has provided precise measurements of the spectro-polarimetric characteristics of planets and brown dwarfs (e.g. HD206893 B \cite{Delorme}; HR2562 B \cite{Konopacky2,Mesa2}; GJ504 b \cite{Kuzuhara}; HD95086 b \cite{Chauvin4}; PDS70 b \cite{Muller}; 51 Eri b \cite{DeRosa1}; HIP65426 b \cite{Cheetham2}) as well as precise measurements of their projected separation \citep[$\sim$ mas][]{Maire3,DeRosa}. Multi-epoch observations of a single system allow for a precise determination of a part of the orbit or, in some cases, for a full characterization of the entire set of orbital parameters of the planet. This was for example the case for the planets of HR8799 \citep{2016A&A...587A..57Z,Wang,GravityCo,Gozdziewski1} and $\beta$ Pic b \citep{Lagrange4}.\\
In this paper we explore the possibility of detecting companions to brown dwarfs (BD) and planets discovered through DI. In fact, the physical characteristics of the latter are particularly suitable for the search of further bound objects (from now on, satellites) in their close vicinity, since they are relatively massive and usually with semi-major axis from tens to hundreds of au. Thus, their satellites can populate a wider area, identified by a large fraction of the Hill radius, around the planet. Similarly to planetary architecture, we may think of using a combination of different methods to explore the surrounding of an exoplanet for the presence of satellites, substituting the star as a reference with the planet itself as observational target. Each detection technique will be then limited to a specific portion of the parameters space.\\
This investigation is stimulated by the arrival of new instrumentation, both from space \citep[e.g.,  MIRI/JWST,][] {Danielski} and ground \citep[e.g PCS/ELT][]{Kasper1}, which will start a new era of the exoplanetary science, as well as by previous studies on the detectability of satellites with different techniques \citep [see e.g.][]{Heller1, Heller2, Vanderburg, Rodenbeck2020, Kipping}. Even if we would be strongly biased towards the detection of massive satellites in the form of giant planets, these instruments might unveil cold worlds that are too faint to be directly detected at present times, including rocky planets and possibly the first exomoons. In addition to this, instruments constructed in the last few years \citep[e.g., GRAVITY,][]{GRAVITY} allow or will soon allow \citep[e.g., CRIRES+ especially with its link to SPHERE AO system, HiRISE;][]{Otten,Vigan3} dedicated characterization studies of directly imaged planets.\\
We notice that the term "moon" is usually dedicated to relatively small rocky bodies which orbit a planet, in analogy with the moons of our solar system. Here, instead, we will investigate a wider mass range of satellites, including objects similar in mass to the central planet, sometimes referred to as binary planets \citep{Podsiadlowski,Ochiai}.\\ Massive triple systems have already been discovered, such as the binary BD system that orbits around $\epsilon$ Indi \citep{King} at a separation of 1500 au with masses in the range $[50,70]$ M\textsubscript{Jup}; the one around GJ 569 B \citep{Femenia} at a separation of $\sim 90$ au and masses in the range $[55,75]$ M\textsubscript{Jup}; and a few other examples such as the pairs of BDs discovered around GJ 417 B \citep{Kirkpatrick}, HD130948 B \citep{Potter} and AB Dor Ca/Cb \citep[via interferometry,][]{Climent}. 
Thus, multiple systems, as exotic as they seem at first glance, are not a remote possibility to be investigated and might extend to smaller masses in the planetary regime.\\
More recently, the first detections of planetary-like satellites were claimed, as in the case of the Neptune-like candidate exomoon detected around the Jupiter-like planet Kepler 1625 B \citep{Teachey}, the  $2.6$ $M_{\oplus}$ candidate satellite for Kepler1708 b \citep{Kipping} and the candidate companion ($\sim 1$ M\textsubscript{Jup}) around the low-mass ($\sim 10$ M\textsubscript{Jup}) brown dwarf DH Tau B \citep{Lazzoni}. Though the existence of these candidates is debated \citep[see e.g. ][]{Rodenbeck, Heller3, Kreidberg} in each case, the ratio between the mass of the satellite and the planet $q_s$ is quite high (0.01, 0.003 and 0.1 respectively) with respect to the solar system's moons which orbit giant planets (where the highest ratio, $q_s=2.37 \times 10^{-4}$, is reached by Saturn's moon Titan). This might open up to a much more variegated spectrum (both in mass, separation and formation mechanisms) of satellites around exoplanets.\\
At the same time, from the properties of the satellites, we might extract useful insights on the formation mechanisms of directly imaged exoplanets and brown dwarfs. In fact, the formation mechanism of the latter is not very clear and likely involves core accretion \citep{Mizuno}, disk instability/fragmentation \citep{Cameron,Boss}, and cloud/filament fragmentation \citep{Larson1985}. Therefore, we can expect that different formation mechanisms deeply shape the properties of putative satellite companions.\\
The purpose of this work, then, is to explore the hunting capabilities of techniques such as RV, transits, astrometry and direct imaging, applied to imaged exoplanets and brown dwarfs. The latter, coupled with the best performing present and future instruments, could ultimately result in the detection of satellites of directly imaged companions.\\
We discuss in Section 2 the scenario for formation of satellites. In Section 3, we analyze the amplitudes of the signals (DI, RV, Transits and Astrometry) generated by a population of satellites around a test planet detected by direct imaging. In Section 4, we provide order of magnitude estimates of the probability of finding satellites to known planets discovered by DI using the various techniques for two different populations of satellites, that might be expected within different scenarios for planet formation. Finally in Section 5, we present a brief summary of the paper.

\section{Exomoons and binary planets}

The satellite category includes a huge variety of objects. In the Solar System, moons of the outer planets are often divided into two distinct classes: the regular and irregular satellites. The regular satellites orbit close to their host planet, moving on prograde orbits, with low inclinations (with respect to the planet's equatorial plane) and  eccentricities \citep{Mosqueira, Mosqueira1, Sasaki}. The irregular satellites, on the contrary, are scattered around their host planets and move on either prograde or retrograde orbits \citep{Nesvorny3, Jewitt1, Holt}, which are often highly inclined and eccentric. These moons are also typically relatively small (with the exception of Triton, the giant satellite of Neptune).

Due to their variety in mass and orbital parameters, multiple models have been proposed for planetary moon formation, and similar mechanisms can be adopted, in principle, to form extrasolar satellites. For the regular satellites of the giant planets, current theories suggest that during the phase of gas accretion, these planets developed their own small sub-nebulae, within which the satellites formed, similarly to the terrestrial planets within the main proto-planetary nebula \citep{Canup2002, Canup2006, Nesvorny4, Mosqueira, Mosqueira1, Mosqueira2, Sasaki}. The irregular satellites, on the other hand, are thought to have been formed elsewhere and then been captured, either through collisions \citep{Goldreich, Woolfson, Koch}, three-body encounters between the host planet and binary planetesimals \citep{Agnor,Vokrouhlicky}, three-body encounters involving two of the giant planets and the captured object \citep{Nesvorny5}, or through gas-drag \citep{Cuk}. Possibly a similar mechanism can be invoked also for the Pluto-Charon system (see e.g. \citealt{Rozner2020}, though the usual explanation for this system is giant impact, \citealt{McKinnon1989,Canup2005}; note, however, that Pluto is a dwarf planet). No mechanism is sufficient to explain all the observed properties of all irregular satellites \citep{Jewitt1}, and so these objects are thought to be representative of the different processes that occurred during the final stages of planetary formation. The Moon is different, since it has formed when the young Earth was involved in a giant collision with a Mars-sized embryo, towards the latter stages of its accretion \citep{Cameron1976,Canup2004,Benz,Reufer}.

All moons in the Solar System have very small masses compared with their planets. Among the regular satellites of the giant planets, the highest value for the satellite mass ratio $q_s$ is that of Saturn to Titan ($q_s=2.37 \times 10^{-4}$). A much higher value of $q_s$ is obtained for the Earth-Moon system ($q_s=0.011$), with an extreme value of $q_s \sim 0.12$ for the Pluto-Charon system; in this last case the barycenter of the system lies out of the main body. 

In relation to the planet formation channels, the lack of giant satellites is expected for a wide variety of planetary systems, being likely a consequence of formation in a core accretion scenario (see e.g. \citealt{Canup2002, Canup2006, Sasaki}). However, planets may possibly form through other mechanisms - such as cloud fragmentation and disk gravitational instability - different from core accretion and for which the formation of giant satellites (or binary planets) is possibly not precluded. For instance, \citet{Inderbitzi2020} considered the formation of satellites in circumplanetary disks around a 10 M\textsubscript{Jup} planet formed by gravitational instability at 50 au from its parent star and found that, under favorable circumstances, a satellite as large as 10 $M_{\oplus}$ might form; this is close to the minimum mass required for runaway accretion. As already mentioned, a few candidate planetary satellites have already been claimed (Kepler 1625 b, Kepler 1708 b and DH Tau B). Somewhat relevant for the present discussion are also the discoveries of close brown dwarf-brown dwarf pairs in wide orbits around stellar hosts \citep[e.g.,][]{King, Femenia, Kirkpatrick, Potter} and of isolated brown dwarf-planet pairs \citep[e.g.,][]{Luhman, Chauvin3}. As we will discuss in next Sections, the detection of giant satellites with $q_s > 0.1$ around massive planets or brown dwarfs appears possibly feasible, though difficult, at present with different techniques. If this were true, the possible discovery of giant satellites will shed light on the mechanisms of formation of their planets. 

\section{Techniques}

To show which effects are produced by satellites orbiting a planet or a brown dwarf, including both the population of \textit{planet-like} and \textit{binary-like} satellites that we will introduce in Section 4, we chose a test system with a directly imaged companion, $\beta$ Pic \citep{Lagrange1}. $\beta$ Pic b is, under many aspects, an instructive candidate for these kind of simulations thanks to its relatively small distance from the Sun \citep[19.4 pc,][]{VanLeeuwen}, its favorable contrast with respect to the central star, the edge-on configuration of the system (which increases the chances of detection with techniques such as RV and transits) and its dense coverage of observations taken in the past decade. The main parameters for $\beta$ Pic b are listed in Table~\ref{t:planet}. The companion is a $12.8$ M\textsubscript{Jup} planet, as obtained by converting its contrast with respect to the central star \citep[$10^{-4}$ at $1.593$ $\mu$m][]{Lazzoni2} with the BEX-COND-warm models \citep{Marleau}, using  an estimated age of $16$ Myrs \citep{Desidera1}. Thanks to observations taken during a time interval spanning over 12 years, the orbit of the planet is well constrained with a semi-major axis of $8.9$ au, an eccentricity of 0.01 \citep{Lagrange4} and a Hill radius of $\sim 1.2$ au. In the next sections we show the results that we obtained with the various detection techniques for a population of satellites injected around $\beta$ Pic b.

The population of satellites that we simulated has masses in the range $[0.001,1]M_P$, where $M_P$ is the mass of the planet, with log-steps of 0.01 M\textsubscript{Jup}. Considering this mass range, we take into account objects starting from the sub-earth mass regime up to binary systems with mass ratio 1. We considered only prograde circular orbits, coplanar with orbital plane of the planet, with semi-major axes in the range $[R_{Roche},R_H/2]$, at log-steps of 0.0005 au, where $R_H$ is the Hill radius of the planet \citep{Raymond2}. The $R_H/2$ limit for stable orbits was adopted following the results published by \cite{Domingos} whereas the inner boundary for stability is determined by the Roche limit, which determines the minimum separation possible for a satellite before the tidal disrupting due to the planet would break it up. The Roche limit is defined by the following equation
\begin{equation}
R_{Roche}=R_{s}\biggl(\frac{2M_P}{m_{s}}\biggr)^{1/3},
\end{equation}
where $R_{s}$ and $m_{s}$ are the radius and mass of the satellite, respectively, and $R_s$ was calculated as described in Section 3.3, using \eqref{eq6}.\\
Note that the detection techniques that we will discuss in the following are applied to the planet $\beta$ Pic b itself, and not to the central star.

\subsection{Direct Imaging}

\begin{figure}
	\includegraphics[width=\columnwidth]{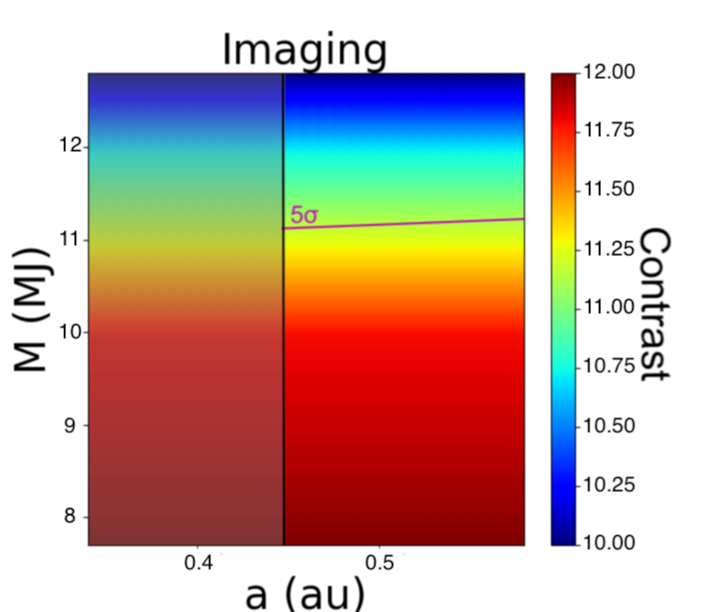}
    \caption{Contrasts with respect to the star related to a population of self-luminous satellites generated around $\beta$ Pic b. The pink line represents the detection limits at 5$\sigma$ level around the planet; the shaded area corresponds to the blind area around the planet within 1/2 FWHM}
    \label{betapic_di}
\end{figure}

We generated a population of satellites with magnitudes in the range $[Mag_P,Mag_P+2]$, where $Mag_P$ is the H-band contrast in magnitude of the planet with respect to the star. Each absolute magnitude was then converted into mass using the age of the system and the BEX-COND-warm models \citep{Marleau}. Results are shown in Figure \ref{betapic_di}, where the shaded region represents the inaccessible area limited by half of the full width at half maximum (FWHM) of the PSF of the planet and contrast are given with respect to the star. Note also that this is the only technique for which we needed to plot the results using linear scales on both axis. In fact, the parameter space for DI is much smaller than the RV, transits or astrometry parameter spaces, for which, instead, we used logarithmic scales..

The 5$\sigma$ level contrast curve shown in pink was derived following the procedure presented in \cite{Lazzoni}, thus carefully subtracting the contribution of the PSF of the planet with dedicated tools and then calculating the standard deviation in concentric annuli with a width of one FWHM each, starting from a separation of 1/2 FWHM from the center. This technique was already applied to the entire sample of substellar objects observed with SPHERE, leading to the discovery of a candidate companion around the low-mass brown dwarf DH Tau B.

In \cite{Lazzoni} we analyzed a collection of 27 substellar companions detected with SPHERE, which is far from being a complete sample. First of all, we had to discard a few systems that were observed in poor weather conditions with resulting low-quality data sets. Also, two of the stars were close visual binaries and we had to exclude those systems from the analysis because we had no proper model PSF for the subtraction. For this kind of situation, an efficient solution could be the Reference Differential Imaging \citep[RDI,][]{Lafreniere}, where a star, similar in magnitude and declination to the parent star of the system, is observed in a range of time very close to the coronagraphic observation \citep[or even during the latter with a technique called star hopping,][]{Wahhaj4}.

Moreover, the sample could be completed with objects that are outside IRDIS/SPHERE FoV and will be reachable with instruments such as the Enhanced Resolution Imager and Spectrograph \citep[ERIS,][]{Kenworthy} which has a FoV of $54"\times54"$ and with targets which are observable only in the Northern hemisphere, using, for example the LBT facility, which will soon be equipped with a high contrast imager in the near infra-red, SHARK-NIR \citep{Farinato}.

Nevertheless, the SPHERE data analyzed so far let emerge the limitation of the technique due to: (i) the strong influence of the presence of the atmosphere, which can cause strong variations of the PSF during the observation and, in turn, unreliable features in the residuals of the object; (ii) and to the contrast reachable, which limit us to Jupiter-like objects. The next generation of instruments, both from ground and space, will overcome these problems. In fact, JWST/MIRI is expected to provide high-contrast imaging in the spectral range [5.6,28.8] $\mu$m, with expected contrasts from $10^{-4}$ to $10^{-6}$ at separations $> 2$ arcsec \citep{Danielski}. Even if the contrast were less deep than what is achievable with SPHERE, JWST/MIRI will be more efficient for colder and smaller objects due to the longer wavelengths coverage and the much better stability of the PSF over time with respect to ground-based observations. This will guarantee the reality of any detectable feature around the companion. Another limitation of this instrument is the dimension of the PSF that increases for increasing wavelengths and limits the detection of satellites at small separations. For example, for a planet like $\beta$ Pic b, at the shortest wavelength available with MIRI, the FWHM of the PSF is $\sim 3.5$ au, which is well beyond the Hill radius of the planet.

On the ground-instrumentation side, METIS on ELT \citep{Brandl} will be able to access science cases similar to the ones presented for SPHERE, even if it is in the mid-infrared. This is due to the fact that the angular resolutions for the two instruments are roughly similar. 
However, with METIS we will have not only a gain in contrast towards less luminous planets but also a much higher quality of the observations. On even larger timescales, a dedicated high-contrast imager such as PCS on ELT \citep{Kasper1} is expected to reach contrasts of $10^{-8}$ at 30 mas and even deeper contrasts ($\sim10^{-9}$) at larger separations. In the near-IR, though, we are limited once more to the detection of massive satellites which, at young ages, still emit at thermal wavelengths. 

\subsection{Radial Velocities}

\begin{figure}
	\includegraphics[width=\columnwidth]{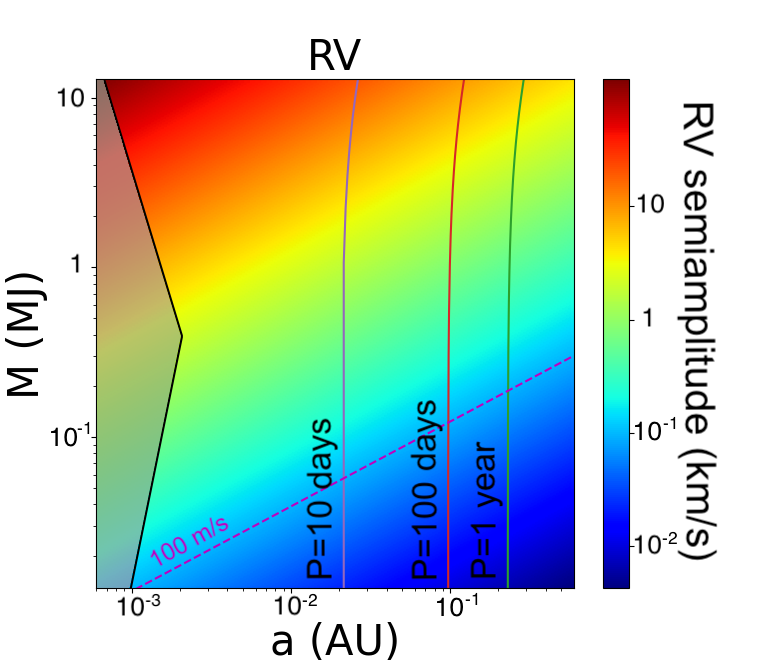}
    \caption{Radial velocity signals related to a population of satellites generated around $\beta$ Pic b. The orbital periods are represented with purple (10 days), red (100 days) and green (1 year) curves and the Roche limits for tidal disruption is shown in black. The pink dashed line represents the detection limits achievable with HiRISE. The shaded grey area corresponds to the tidal disruption region.}
    \label{betapic_rv}
\end{figure}

The amplitude of the motion of the planet in the radial direction due to the presence of a satellite was calculated using the equation given by \cite{Wright2}, substituting the parameters of the star with the parameters of the planet, and the ones of the planet in the original expression with the parameters of the satellite. We thus obtain
\begin{equation}
K=\frac{m_s \sin{i_s}}{M_P}\sqrt{ \frac{GM_P}{a_s}}\frac{km}{s*1000}
\end{equation}
where $m_s$ and $a_s$ are the mass and the semi-major axis of the satellite and $i_s$ is the inclination of the satellite with respect to the plane of the planet. For the following discussion, we will consider the satellite being co-planar with the planet ($i=90^\circ$).

As shown in Figure \ref{betapic_rv}, RV amplitudes of a few hundred of m/s are already reachable for the smallest satellites with masses $\lesssim 0.05$ M\textsubscript{Jup}, if placed within 0.01 AU. We note, however, that this is an optimistic estimation given the favorable configuration considered. If the planet-satellite system is not aligned, the RV signal generated differs significantly from case to case. For example, a co-planar satellite of $0.05$ M\textsubscript{Jup} at 0.01 au gives an RV signal of $0.14$ km/s whereas the same object on an inclined orbit of $20^\circ$ produces $K=0.05$ km/s, which would be not detectable.

The measurement of the RV of a directly imaged planet first has to face the issue of properly isolating the light of the planet itself from the bright halo of the central star. Therefore, it is mandatory to couple the spectrograph with an AO module. First efforts were already performed on some single-epoch measurements taken with CRIRES/VLT of $\beta$ Pic b \citep{Snellen} and GQ Lup B \citep{Schwarz2016} and measurements taken with NIRSPEC/Keck of DH Tau B \citep{Xuan} and HR8799 b and c \citep{Ruffio}. From these first tentative characterizations, it emerged how the detection of variations in RV signals of planets is not only strongly limited by the low photon-noise ratio but also by the contamination of the central star.

The intrinsic variability of planetary RV signals due, for example, to the presence of clouds in the atmosphere, to the accretion of material from the surrounding environment or to the orbit of the planet around the star, should be taken into account as well in such analyses and possibly identified through multi-epoch observations.

\cite{Vanderburg} presented a detailed description of the RV signal generated by the candidate exomoon of Kepler-1625 b and detection feasibility of a similar exomoon around directly imaged planets with current and future instruments. They consider, besides the Doppler signal due to orbital motion of the satellite around the planet and the planet around the star, the effects of partial planet illumination, of activity or heterogeneous cloud coverage, the light contribution of the exomoon with respect to the planet (peak-pulling), and the partial occultation by disk clumps. Therefore, we do not repeat here their evaluations and rather focus our attention on improvements thanks to forthcoming instruments.

In fact, relevant advances with respect to the early single-epoch attempts mentioned above can be obtained with updated instruments as CRIRES+ \citep{Follert} and even more when this is coupled with the extreme-AO system of SPHERE as conceived with HiRISE \citep{Otten,Vigan3}. As shown in Figure 9 of \cite{Vigan4}, with HiRISE we expect contamination as small as a few $10^{-4}$ for a planet like $\beta$ Pic b. 

To estimate the typical accuracy achievable in the measures of a planet RV signal with an instrument such as CRIRES+, we considered the following steps. Using the Exposure Time Calculator (ETC: https://etc.eso.org/observing/etc/crires) the expected signal-to-noise ratio (SNR) in the H-band on $\beta$~Pic b with CRIRES+ is about 40 per pixel in 1 hr (with $6\times 600$~sec DITs) using the natural guide star adaptive optics (AO) on the primary. However, in this mode the dominant source of noise for $\beta$~Pic~b should be the stellar background. 
The latter is efficiently removed using HiRISE, which, however, is not currently included in the ETC. It should be noted that HiRISE will be less transmitting than the direct train to CRIRES+ (with the AO) by a factor of about 15 - though this is partly offset by the three times higher Strehl provided by the Sphere Ao for eXoplanet Observation \citep[SAXO][]{Fusco1} of SPHERE with respect to Multi-Application Curvature Adaptive Optics \citep[MACAO][]{Paufique} available for CRIRES+.

We simulate this lower efficiency considering an object with H=15.5, rather than H=13.8. The SNR per pixel in this condition is about 10. To transform this into an error in RV, we may consider the observations of the late M-star TW Hya made by \citet{Figueira2010} where an internal accuracy of 6 m/s was obtained from spectra with SNR$\sim$230 (this value has been traced back using the ETC, considering the stellar magnitude, the observing mode, DIT and number of exposures). Ignoring the much larger wavelength coverage of CRIRES+ with respect to CRIRES, the expected RV accuracy depends on the inverse of SNR and then observations with a SNR=10 for a late M-object (as $\beta$ Pic b) should then be $\sim 140$~m/s, but it can well be a factor of 1.5 better considering the much wider spectral range. We then assume an error of 85 m/s in 1 hr observation on $\beta$ Pic b. Incidentally,
if the lines of the planet and the satellite are not blended, detection of the RV variations from the satellite will provide a direct determination of the mass of the planet, crucial in modelling young objects.

We can then set a maximum detectable amplitude at $\sim 100$ m/s (highlighted by a pink dashed line in Figure \ref{betapic_rv}). 
We note that this threshold is based on the analysis with CRIRES of very late type stars (M5 and later), with spectra similar to those of bright DI exoplanets/BDs considered in this paper. Thus, on the one hand, the threshold obtained could result in an underestimate of detection capabilities, given the larger spectral coverage of CRIRES+. On the other hand, some substellar companions in the sample here have L/T spectral types, implying that the radial velocity signal is different from the one actually adopted.
Even if the maximum detectable amplitude of $\sim 100$ m/s is quite optimistic for actual instrumentation, it is quite realistic for HiRISE. In the $\beta$ Pic b system this limit, together with the orbital period of the injected object around the planet, strongly constrains the characteristics of detectable satellites. Periods from 10 days to 1 year are plotted as vertical lines. At wider semi-major axis the orbital period increases, and, in turn, the probability of a detection drastically decreases.

A possible caveat to this technique, which is shared by transits and astrometry as well, is the time series of observation needed for the detection of a satellite. To improve the detection limits, several observations with top-class instrumentation on 8m telescopes are needed and this could be significantly time consuming, limiting the number of substellar objects that could be investigated.

\subsection{Transits}

\begin{figure}
	\includegraphics[width=1.11\columnwidth]{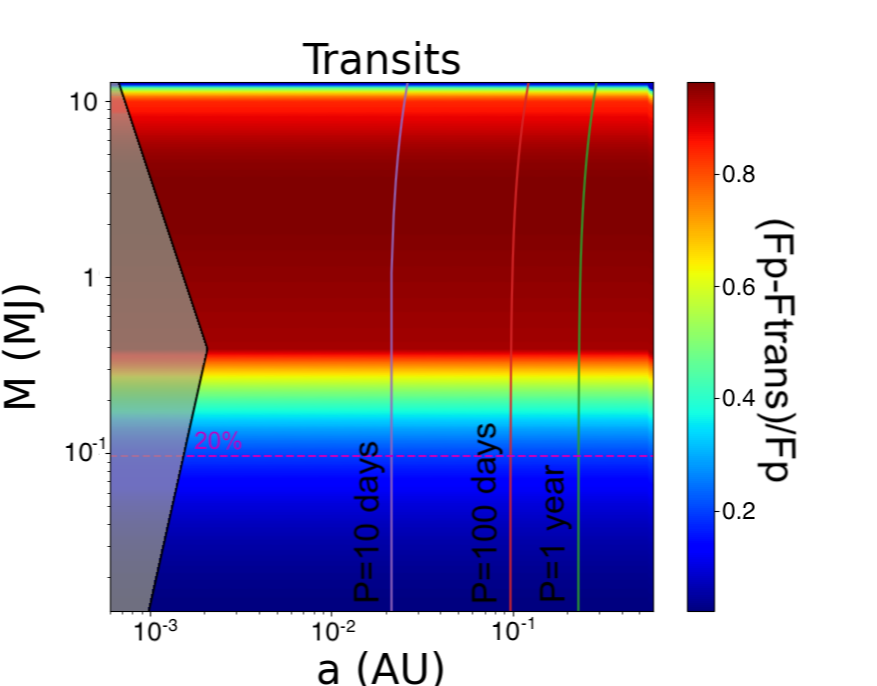}
    \caption{Transit signals related to a population of satellites generated around $\beta$ Pic b. The orbital periods are represented with purple (10 days), red (100 days) and green (1 year) curves and the Roche limits for tidal disruption is shown in black. The pink dashed line represents the detection limits achievable with SPHERE. The shaded grey area corresponds to the tidal disruption region.}
    \label{betabic_trans}
\end{figure}

As mentioned above, the masses of the population of satellites generated around $\beta$ Pic b vary from one-thousandth of the mass of the planet, thus $\sim 0.012$ M\textsubscript{Jup}$\sim 3.8$ $M_{\oplus}$, to the mass of the planet itself. From both theoretical and observational studies we know that the radius of a planetary object tends to increase with its mass up to a limiting value \citep[here we adopt $0.39$ M\textsubscript{Jup} following][]{Bashi}. For larger masses, instead, the radius settles roughly at 1 R\textsubscript{Jup}, with a much slower increasing rate with respect to the mass. Since there is yet no study providing a mass {\it vs.} radius relation for satellites (either rocky or giants ones), we will adopt for the latter Eq \eqref{eq6}. This is likely a good approximation because of the similar steps which take to the formation of both satellites (either within the circumplanetary disk or within the circumstellar disk and subsequent capture) and planets. The interactions between the primary object and the satellite (e.g. inflation of the latter due to heating caused by tidal locking) might cause deviations from the mass {\it vs.} radius relation adopted for satellites, but such an accurate analysis is beyond the general purposes of this paper.

Following the previous discussion, giant satellites are expected to have radii which are comparable with the radius of $\beta$ Pic b itself. Thus, these objects could potentially completely obscure the primary planet. However, with increasing mass, a non-negligible contribution to the observed flux might come from the satellite itself. In fact, young giants are expected to be self-luminous so that a transiting massive satellite will cause a flux modulation more than an eclipsing transit, similarly to a binary star flux modulation.

In order to take into account these effects, we first consider the satellite as an obscurer, calculating the dimming in the flux following the equation \citep{Seager} 
\begin{equation}
\frac{F_P-F_{trans}}{F_P}=\frac{R_s^2}{R_P^2}
\end{equation}
where $R_P$ and $R_s$ are the planet and satellite radii, respectively. As mentioned before, the radii are calculated with the relations given by \cite{Bashi}\footnote{This mass-radius relation was derived for systems of different ages. For the very young Jovian planets that we are considering we should expect a larger radius since gravitational contraction is still ongoing.} 
\begin{equation}
R=
\begin{cases} 
129362\cdot\bigl(\frac{m_s}{M\textsubscript{Jup}}\bigr)^{0.55} km       & \text{if  $m_s<0.39M\textsubscript{Jup}$ }\\
77818\cdot \bigl(\frac{m_s}{M\textsubscript{Jup}}\bigr)^{0.01}km	 & \text{if  $ m_s>0.39M\textsubscript{Jup}$}
\end{cases}
\label{eq6}
\end{equation}
Then, the contribution to the total flux given by the satellite  is added to take into account its self-luminosity. This is done, only for satellites more massive than $0.2$ M\textsubscript{Jup} because the BEX-COND-warm models \citep{Marleau}, considered for the mass-luminosity conversion, are not available  below that value \citep{Linder}. This is not a drastic issue since the self-luminosity of planets drastically decreases with the mass and the contribution to the flux below $0.5$ M\textsubscript{Jup} is negligible.

We show in Figure \ref{betabic_trans} the results obtained for the population of transiting satellites injected around $\beta$ Pic b. Given the nearly edge-on configuration of the latter ($i=89^{\circ}$), the injected population of satellites, coplanar with the orbital plane of the $\beta$ Pic b, will transit with good approximation through the center of the planet. We will discuss at the end of this Section the influence of the inclination on the detectability of satellites.

As shown in Figure \ref{betabic_trans}, the limit for a detectable occultation/modulation for an instrument provided of an AO system is set at a $20\%$ diminishing of the flux of the planet \citep[following the discussion presented in][]{Biller2}. This is justified by a combination of photometric precision reachable by AO instruments, which are mainly limited by variation of the observational parameters, and intrinsic variability due, for example, to the presence of clouds in the atmosphere. This implies that putative satellites around $\beta$ Pic b with masses $<0.1$ M\textsubscript{Jup} would not be detectable. Also, there is an inverse trend for massive satellites, since the incoming flux of the latter for objects with $m_s\sim M_P$ can reach the flux of the planet itself. In these cases, the two fluxes are very much alike and the modulation is not detectable. However, we note that this trend is present only for the very upper part of the mass range tested since the flux rapidly decreases with the mass. We also note that the already unfavorable limit of 20 $\%$ would increase further for non-AO observations, making the transit technique somewhat unattractive for the hunting of satellites around DI companions. This is also showcased by the very low probability ($p_i<2\%$, see Table \ref{t:prob1} and \ref{t:prob2}) of detecting a satellite in the sample of 38 substellar objects considered, as obtained in Section 4. 

A first study of photometric variation was applied to HR8799 b and c \citep{Biller2} using SPHERE observations. Taking into account the inclination of the system, $\sim 27^{\circ}$, the derived sensitivity for the two planets is $11-22\%$ and $>50\%$, respectively. In the same paper, it is also estimated the signal that would be generated by a population of putative transiting satellites injected around HR8799 b, considering the system as seen edge-on.

Photometric variation of a planet can be associated to variable features in the atmospheres, such as clouds. To disentangle the presence of a satellite from the intrinsic mutations of a cloudy atmosphere we could perform multiple observations of the planet at different wavelengths. In fact, if a satellite exhibits, a diminishing in the flux (or a modulation of the latter) it is expected to be almost achromatic and periodic whereas clouds should appear only at certain wavelengths and variable due to their heterogeneous distribution.

Photometric observations of isolated planetary mass objects \citep{Limbach2021} would represent a useful pathway to understand their intrinsic variability and to demonstrate the presence of satellites around these objects. The comparison of the frequency of exomoons around isolated planetary-mass objects with those around planetary companions orbiting stars would also be highly interesting for our understanding of their formation mechanisms.

Even if the orbital separation is not straightforwardly included in the calculation of a transit, the cadence with which the object passes in front of the planet is fundamental to confirming the presence of the former. Continuous or quasi-continuous long-term variability monitoring of directly-imaged planets appears prohibitive in terms of observing time requirements (considering the high-quality AO and then large aperture size needed). Thus, satellites with a short period ($\le 10$ days) are more likely to be revealed, limiting in turn the range of detectable satellites around $\beta$ Pic b.

\begin{figure}
\centering
\includegraphics[width=1.11\columnwidth]{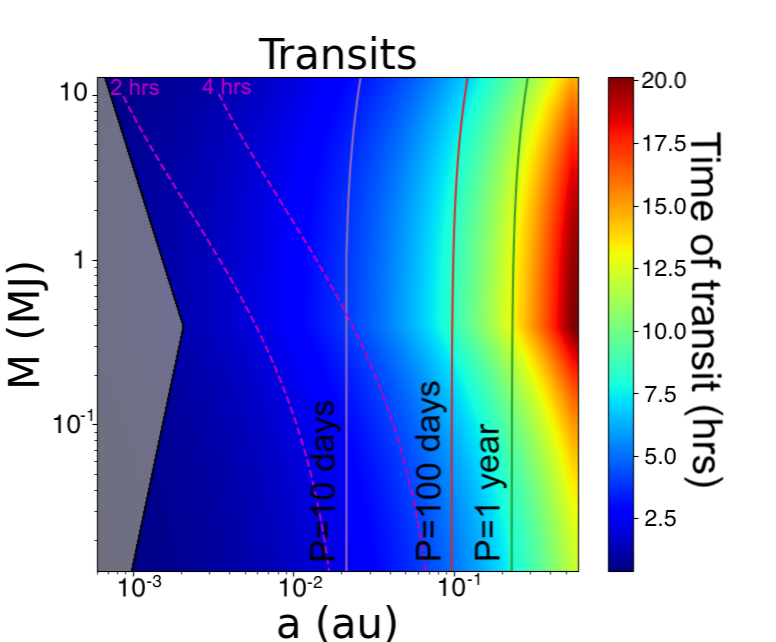}
\caption{Duration of transits of a population of satellites injected around $\beta$ Pic b. The orbital periods are represented with purple (10 days), red (100 days) and green (1 year) curves and the Roche limits for tidal disruption is shown in black. The pink dashed line represents the upper limits to the observable time of the transit. The shaded grey area corresponds to the tidal disruption region.}
\label{betapic_tt}
\end{figure}

Another interesting parameter from an observational point of view is the duration of the transit itself given by \citep{Seager}
\begin{equation}
T_{tr}=\sqrt{\frac{4a_s^3}{G(m_s+M_P)}}\sin{^{-1}\biggl(\frac{\sqrt{(R_P+R_s)^2-(bR_P)^2}}{a_s}\biggr)},
\label{eqttr}
\end{equation}
where $b$ is the impact parameter, i.e. the distance from the center of the planet at which the satellite transits.
In Figure \ref{betapic_tt} an additional plot regarding the transit technique is shown, with the duration of the transit in place of the dimming in the flux (in this case we considered only satellites transiting through the center of the planet, $b=0$). Considering an instrument from the ground, such as SPHERE, we set an upper limit to the observable time of the transit between 2 and 4 hours. Above this value, variable weather conditions and the increasing distance from meridian passage may strongly influence the observations and the speckles subtraction performed during post-processing (e.g. ADI, PCA). With this further constraint, the zoo of detectable satellites strongly decreases. One possible way to detect satellites with long lasting transits could be to observe the target in different nights with shorter sequences. However, in this case, the photometric precision of the instrument might not be negligible \citep[e.g. the rms scatter of contrast from multi-epoch observations of reference systems with  SPHERE was estimated to be typically 0.05-0.08 mag in H band,][]{Langlois2}.

The inclination of the planet-satellite system is a crucial parameter to take into account when considering transit observations. As mentioned before, for this test case we considered satellites being coplanar with the orbital plane of the planet. This likely includes objects generated within the circumplanetary disk \citep[regular satellites, ][]{Mosqueira,Mosqueira1,Sasaki}. The irregular satellites \citep{Nesvorny3,Jewitt1,Holt}, which are usually the outcome of capture from the surrounding environment, are expected to orbit on inclined orbits. Since, to have a transit, the geometrical condition $cos{i_s}\le(R_P+R_s)/a_s$ must be respected, irregular satellites with high values of $i_s$ will pass in front of the parent planet with less probability.

Similar arguments apply to systems which are seen further from edge-on. HR8799 planets, for example, are inclined at $\sim 40^{\circ}$ with respect to the line of sight \citep{Booth2} and any regular satellite would be impossible to detect. Let us consider, for example, the outermost planet, HR8799 b, which has a mass of 6.8 M\textsubscript{Jup} \citep{Lazzoni3} and thus a radius  $R_P=1.13$R\textsubscript{Jup}. If we assume that regular satellites were formed from the circumplanetary disk, we can consider terrestrial-like objects to orbit around HR8799 b. A 1 $M_{\oplus}$ should orbit at a maximum distance of $7\times10^{-4}$ au to transit in front of the planet, which is at the very limit of disruption due to tidal forces ($6.9\times10^{-4}$ au).

It is worth mentioning, however, that the inclination could also have a positive impact on the detectability of a putative satellite since it could significantly decrease the time of the transit, as shown in Equation \eqref{eqttr}.

\subsection{Astrometry}

\begin{figure}
	\includegraphics[width=\columnwidth]{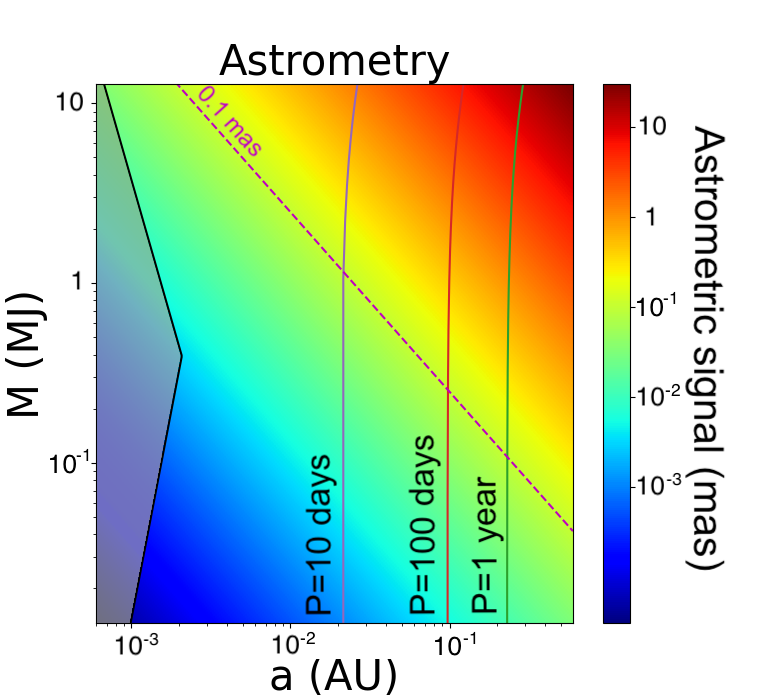}
    \caption{Astrometric signals related to a population of satellites generated around $\beta$ Pic b. The orbital periods are represented with purple (10 days), red (100 days), green (1 year) and orange (10 years) curves and the Roche limits for tidal disruption is shown in black. The pink dashed line represents the detection limits achievable with GRAVITY. The shaded grey area corresponds to the tidal disruption region.}
    \label{betapic_astr}
\end{figure}

The deviation of the orbit of the planet in the transverse direction due to the presence of the satellite was calculated using
\begin{equation}
\theta=\frac{m_s}{M_P}\frac{a_s}{d} arcsec
\end{equation}
where $d$ is the distance of the system expressed in parsec. The astrometric measure we consider here is the relative position of the planet with respect to the star; the presence of the satellite is deduced from the residuals of the planet motion with respect to a simple Keplerian orbit around the star. If the planet orbit is much longer than the satellite one, then detection of the wobble due to the satellite does not require to follow the full planet orbit, but only the full satellite orbit. For this reason we set an upper limit of 20 yr to the period of orbits detectable through astrometric signal.

In principle, we cannot tell from this data if the wobbling is due to a satellite around the planet or an additional planet around the star; however, absolute astrometry from Gaia may solve this ambiguity in the near future. Thanks to very precise astrometric measurements performed by instruments such as SPHERE, we are able to detect deviations down to 1 mas \citep{Maire5}. Even smaller deviations (potentially down to 100 $\mu$as, pink dashed line in Fig. \ref{betapic_astr}; see e.g. \citet{Lacour2021}) are achievable with instruments such as GRAVITY/VLTI \citep{GRAVITY} for bright targets ($m_K\sim 10$~mag)\footnote{Since we are interested here to the relative motion of the planet with respect the star, limiting magnitude of the target we have to consider here is that of the star}. Thus, for $\beta$ Pic b, the smallest satellite that we should be able to detect if placed at the Hill radius of the planet could have a mass of $\sim 0.02$ M \textsubscript{Jup}. If we get closer to the planet, the mass of detectable objects significantly increases.

When we consider this technique we have, nevertheless, to take into account other factors. First of all, the better we know the orbit of the planet and the better we can appreciate small deviations from its path. This is, for example, the case of  $\beta$ Pic b that has been observed for more than a decade and, given its relatively small semi-major axis and moderately short orbital period, all the orbital parameters are well constrained. Instead, for most of the planets and brown dwarfs detected with the direct imaging technique only a small portion of the trajectory is known and a family of orbits is considered as possible solutions \citep{Blunt, Maire2}. In these cases, satellites become detectable only if the astrometric signals generated are not comparable in terms of amplitude and frequency with the motion of the planet around the star. Also, astrometric variations in terms of relative separation between the star and the planet can be due to the presence of further planetary companions around the star. The $\beta$ Pic system, in this sense, is a very instructive test-bench since it hosts a further planet. $\beta$ Pic c \citep{Lagrange5} was detected via RV technique so that we can marginally break this degeneracy and isolate its contribution to the relative motion of $\beta$ Pic b. Assuming, for $\beta$ Pic c a separation of 2.7 au and a mass of $7.8$ M\textsubscript{Jup} \citep{Lagrange5}, the astrometric signal produced by the planet on the parent star is of $0.59$ mas. Considering a revolution period for the planet of $3.4$ yrs, we can then compare this astrometric signal with the same one, in terms of amplitude and orbital period, that could be produced by a hypothetical satellite around $\beta$ Pic b. As shown in Figure \ref{betapic_astr}, a satellite of $\sim 0.3$ M\textsubscript{Jup} on a 0.5 au orbit with a period of $3.4$ yrs would produce on the b planet a similar astrometric signal as the inner detected planet. Thanks to the detection of $\beta$ Pic c via radial velocities, we can discard the presence of this kind of satellites. This is only an example of the kind of ambiguity that may arise while considering only the variation in relative separation between the central star and the planet.

Therefore, it stands clear that the association of astrometric variations to the orbit of the planet and the presence of putative satellites around the latter becomes feasible not only when the orbital parameters of the planet are well constrained but also when the system is investigated with other techniques. In fact, the detection/non detection of further planets closer to the star via RV and transits and the absolute astrometry of the star given by GAIA \citep{GAIA, GAIA1} are necessary tools to determine the relative astrometry of the planet and to infer the presence of any satellites bound to it.

\section{ Probability of planet detections using the different methods}

In this Section we explore the chances of detecting satellites to directly imaged brown dwarfs and planets, in the range of separation from the star between 0.1 and 5 arcsec, that roughly corresponds to semi-major axis in the range from a few to a few hundreds of au.

Hereinafter we will make a first exploration of this field, considering the feasibility of programs that aim to detect satellites around the directly imaged planets using various observation techniques. To this purpose we will first construct a sample of known exoplanets and then consider two different parametric satellite populations around them. For each technique, we will make a first order of magnitude estimate of the detection probability for each of the two satellite populations. We notice that the assumptions we make (e.g. circular orbits; satellites observed in quadrature) are very rough and simplified so that they should be revised for an appropriate interpretation of any specific survey devoted to satellite detection, and that the predictions on the detection probability we give should only be considered as order of magnitude estimates.

\subsection{Exoplanet sample}

Table~\ref{t:planet} contains the main parameters for the sample of exoplanets and brown dwarfs considered throughout our discussion, along with the references from which they were obtained. While quite large, the sample does not have the aim to be complete, but rather to be representative of those systems known at present for which we may reasonably expect satellite detection. We considered 37 planets in 32 systems. We only considered objects with separation in the range between 100 and 5500 mas, because objects at shorter separation are very difficult to observe with the accuracy required to detect satellites with the techniques described in this paper. Moreover, planets and brown dwarfs at such separations have uncertain formation origins and the detection of any satellite around them could give insightful hints on their history. The parameters reported are only those that are used in our probability estimates.

The planet K-band magnitude used in this paper are also given in Table~\ref{t:planet}. Note that they are derived with the formulas considered in this paper - starting from masses and ages - and they are not the observed values.

\begin{table*}
    \centering
    \caption{Planets and brown dwarfs parameters}
    \begin{tabular}{lccccccccl}
\hline
Planet Name	&	Age	&	Parallax	&	$K_*$	&	$K_p$	&	Sep	&	a	&	$M_*$	&	$M_p$	&	Ref	\\
	&	Myr	&	mas	&	mag	&	mag	&	mas	&	au	&	M$_\odot$	&	M$_J$	& \\
\hline
1RXS J160929.1-210524 b	&	10.0	&	7.16	&	8.916	& 16.9 &	2,215.0	&	309.4	&	0.85	&	8.0	& \citet{Lafreniere2008, Lafreniere2010} \\
2M1207 b	&	8.0	&	19.08	&	11.945	&	15.6 &878.0	&	42.0	&	0.03	&	5.0	& \citet{Chauvin2004} \\
51 Eri b	&	24.0	&	33.58	&	4.537	& $>$21 &	434.0	&	11.2	&	1.75	&	3.6	& \citet{DeRosa2020} \\
AB Pic b	&	45.0	&	19.95	&	6.981	& 15.1 &	5,400.0	&	270.6	&	0.86	&	14.0	& \citet{Langlois2021} \\
beta Pic b	&	16.0	&	51.44	&	3.480	& 14.9 &	510.8	&	8.9	&	1.64	&	12.8	& \citet{Desidera1,Lagrange4} \\
CT Cha b	&	1.4	&	5.21	&	8.661	& 14.8 &	2,680.0	&	514.0	&	0.80	&	15.0	& \citet{Sheehan2019} \\
DH Tau B	&	1.4	&	7.39	&	8.824	& 14.7 &	2,350.0	&	318.1	&	0.10	&	10.6	& \citet{Lazzoni, Sheehan2019} \\
eta Tel B	&	24.0	&	21.11	&	5.010	& 13.2 &	4,210.0	&	199.4	&	2.18	&	47.0	& \citet{Langlois2021} \\
GJ504 b	&	4000.0	&	56.86	&	4.033	& $>$21 &	2,490.0	&	43.8	&	1.18	&	23.0	& \citet{Bonnefoy2018} \\
GQ Lup b	&	3.5	&	6.59	&	7.096	& 13.5 &	712.0	&	117.0	&	1.03	&	30.0	& \citet{Stolker2021} \\
HD1160 c	&	50.0	&	7.94	&	7.040	& 14.2 &	773.0	&	97.3	&	2.00	&	66.0	& \citet{Mesa2020} \\
HD4747 B	&	2300.0	&	51.95	&	5.305	& 13.6 &	590.0	&	11.4	&	0.86	&	70.0	& \citet{Peretti2019} \\
HD19467 B	&	8000.0	&	31.22	&	5.401	& 14.2 &	1,631.0	&	44.1	&	0.95	&	74.0	& \citet{Maire2020a} \\
HD72946 B	&	1600.0	&	38.65	&	5.467	& 13.5 &	235.0	&	6.5	&	0.99	&	72.4	& \citet{Maire2020b}	\\
HD95086 b	&	12.0	&	11.57	&	6.789	& $>21$ &	630.1	&	58.0	&	1.60	&	4.5	& \citet{Desgrange} \\
HR2562 B	&	750.0	&	29.41	&	5.020	& 17.4 &	640.0	&	21.8	&	1.37	&	29.0	& \citet{Sutlieff2021} \\
HR3549 B	&	125.0	&	10.49	&	6.044	& 16.8 & 	850.0	&	81.1	&	2.00	&	48.0	& \citet{Mesa2016} \\
HR8799 b	&	42.0	&	24.22	&	5.240	& 20.6 &	1,720.6	&	72.2	&	1.52	&	5.8	& Zurlo et al. 2022, accepted \\
HR8799 c	&	42.0	&	24.22	&	5.240	& 18.2 &	955.1	&	41.6	&	1.52	&	7.6	& Zurlo et al. 2022, accepted \\
HR8799 d	&	42.0	&	24.22	&	5.240	& 16.9 &	689.8	&	26.9	&	1.52	&	9.2	& Zurlo et al. 2022, accepted \\
HR8799 e	&	42.0	&	24.22	&	5.240	& 18.2 &	397.2	&	16.3	&	1.52	&	7.6	& Zurlo et al. 2022, accepted \\
HIP64892 B	&	16.0	&	7.99	&	6.832	& 16.3 &	1,270.5	&	159.1	&	2.35	&	33.0	& \citet{Cheetham2018} \\
HIP65426 b	&	14.0	&	9.16	&	6.771	& $>21$ & 	824.0	&	115.0	&	1.96	&	8.0	& \citet{Chauvin2017} \\
HIP74865 B	&	15.0	&	8.10	&	7.808	& 15.7 &	201.0	&	24.8	&	1.72	&	46.0	& \citet{Hinkley2015} \\
HIP78530 B	&	11.0	&	7.28	&	6.903	& 18.4 &	4,180.0	&	573.8	&	2.75	&	20.0	& \citet{Langlois2021} \\
HIP79098 B	&	10.0	&	6.83	&	5.707	& 14.7 &	2,359.0	&	345.2	&	4.00	&	20.0	& \citet{Janson2019} \\
HIP107412 B	&	700.0	&	24.51	&	5.593	& 17.6 &	252.2	&	12.8	&	1.32	&	25.0	& \citet{Romero2021} \\
k And b	&	47.0	&	20.00	&	4.320	& 13.9 &	875.6	&	103.6	&	2.70	&	20.0	& \citet{Carson2013} \\
PDS 70 b	&	5.4	&	8.82	&	8.542	& 15.2 &	173.5	&	20.1	&	0.98	&	7.9	& \citet{Wang2021} \\
PDS 70 c	&	5.4	&	8.82	&	8.542	& 15.2 & 	213.2	&	33.2	&	0.98	&	7.8	& \citet{Wang2021} \\
PZ Tel B	&	24.0	&	21.22	&	6.366	& 13.0 & 	558.0	&	70.9	&	0.90	&	52.0	& \citet{Maire2016}	\\
TYC 7084-794-1 B	&	140.0	&	44.63	&	7.046	& 15.3 &	2,990.0	&	67.0	&	0.50	&	32.0	& \citet{Langlois2021} \\
TYC 8047-232-1 B	&	42.0	&	11.59	&	8.405	& 16.4 &	3,210.0	&	277.0	&	0.82	&	13.8	& \citet{Langlois2021} \\
TYC 8998-760-1 b	&	17.0	&	10.57	&	8.392	& 18.2 &	1,712.5	&	162.0	&	1.00	&	14.0	& \citet{Bohn2020}	\\
TYC 8998-760-1 c	&	17.0	&	10.57	&	8.392	& $>21$ &	3,373.0	&	320.0	&	1.00	&	6.0	& \citet{Bohn2020} \\
TYC 8984-2245-1 b	&	13.9	&	9.09	&	8.358	& $>21$ &	1,050.0	&	115.0	&	1.10	&	6.3	& \citet{Bohn2021} \\
GSC 6214-210 B	&	10.0	&	9.19	&	9.152	& 14.8 &	2,205.1	&	240.0	&	0.90	&	14.0	& \citet{Pearce2019} \\
\hline
\end{tabular}
\label{t:planet}
\end{table*}

\subsection{Satellite populations}

The probability of a satellite to be detected with a certain technique must be joined with the probability that the same satellite formed with suitable properties to be actually observed. To explore a little further these points, we considered in this paper two different populations of satellites. For simplicity, we will assume that their properties are defined by parametric distributions. Given that we know very little about satellites of exoplanets, we considered two very different populations. Following the approach described in \citet{Vigan2021}, we assumed that these two populations are drawn from parametric distributions as a function of the mass ratio $q_s$ between the components (where $q_s \leq 1$) and on the orbital semimajor axis $a$. In addition, we considered a random on-sky distribution of the satellite orbit inclination $i$, that is, we will ignore what we know about the planet orbit inclination and the chances that the satellite orbit is coplanar with the planet orbit. The reason for this choice is that for the vast majority of the exoplanets discovered through direct imaging the orbital inclination is very poorly constrained. A significant exception is $\beta$~Pic~b \citep[][]{Lagrange4}, that is our archetype.

(i) In analogy with the approach of \citet{Vigan2021}, for the first case, we are inspired by typical populations of exoplanets that are thought to form mostly by core accretion; we will then call this population {\it planet-like}. For the orbital distribution of gas giant planets, we assume a Gaussian distribution in $\log{a}$, $a$ being the semi-major axis, with fixed mean and $\sigma$. These properties likely depend on the host star mass and based on results to date, we adopt a log-normal distribution with a $\sigma = 0.52$, as used for M stars \citep{Meyer2018,Fernandes2019}. For the mean value, instead, we used the expression $a=\log_{10}(0.45*(M_P/30.0)^{0.77})$, where $M_P$ is the mass of the planet expressed in Jupiter masses. This formula was obtained by assuming that the peak of the satellite formation is close to the ice-line around very young planets. This position was then derived by fitting the isochrones at 1 Myr from \citet{Baraffe2003}. Since this population is thought to form through core accretion within a circumplanetary disk, the relevant typical value for $a$ is fixed here in relation to the snow-line. This depends on the planet luminosity which, in turn, depends on the mass and age - at the epoch of formation of the satellite (that is not the current age of the system). For reference we considered an age of 1 Myr and derived the luminosity using the DUSTY-AMES isochrones by \citet{Allard2001}. As to the planet mass function, we assume a power-law where the frequency $f$ depends on the ratio of the satellite to the planet mass, $q_s = M_s/M_p$, that is $f \propto q_s^{\beta}$, with $\beta = -1.31$ \citep{Cumming2008,Wagner2019}. To estimate the probability of finding such satellites, we may assume that 10$\%$ of the planets have a satellite with $q_s>0.0003$, similar to the frequency of giant planets \citep{Cumming2008}. To place these values in a context, the Titan/Saturn system has a mass ratio equal to $q_s=0.00024$ and Titan is approximately at 60$\%$ of the peak of the separation distribution we assumed. Hence, Titan would be quite a typical satellite extracted from such a distribution. 

(ii) The second population of satellites we considered is under the assumption that planet-satellite systems are analogues to binary systems. We call this population {\it binary-like}. Given the typical range of separations from the parent planet/BD, this population might have formed through fragmentation/instability of the disk around it - likely very early in the formation phase of this system -, fragmentation of the cloud/filament from which the star formed, or by captures \citep{Ochiai}.  For similarity with the stellar binary case we assumed a log-normal semi-major axis distribution of binary companions, as measured for stellar masses (e.g. \citealt{Raghavan2010} for FGK stars and \citealt{Winters2019} for M dwarfs) with mean $\log a = 1.30$ and $\sigma = 1.16$. For this population, we assume a companion mass ratio distribution, which is roughly flat with the mass ratio (power-law slope of 0.25; \citealt{Reggiani2013}) with a minimum mass ratio of $q_s=0.01$. When estimating the probability of finding similar satellites, we may assume that 20$\%$ of the planets have such a satellite that is not far from the typical fraction of binaries among small mass stars \citep{Moe2017}. As a reference, the BD binary companions to $\epsilon$~Ind have a mass ratio $q\sim 0.6$ and are at a projected separation of 26 au from each other \citep{McCaughrean2004} at about 1500 au from the star, that is around 0.05 times the Hill radius. So they may be considered representative of this population. 

In both cases we assume that the satellites should lie within the allowed range of semimajor axis between the Roche limit and 0.5 times the Hill radius.

\subsection{Detection limits}

\subsubsection{Direct Imaging}

We assume that the limiting satellite star contrast depends on three factors: (i) the noise associated to sky background, $c_s$ (this is a term that only depends on the satellite apparent magnitude); (ii) the noise associated to primary background, $c_*$(depends on the K magnitude of the primary and on star-planet separation, neglecting the difference between planet and satellite separation); and (iii) the noise associated to  planet background, $c_p$ (depends on the K magnitude of the planet and on planet-satellite separation).
The following calculations are obtained assuming good-quality observations of 1 hr taken from a 8-meter size telescope during meridian passage.

We may write the limiting contrast as
\begin{equation}
c_{\rm lim} = -2.5 \log{\sqrt{10^{-0.8 c_{\rm s}} + 10^{-0.8 c{\rm_*}} + 10^{-0.8 c_{\rm p}}}}
\end{equation}
where the contrasts are in magnitudes.

The sky background limiting contrast in the K-band is
\begin{equation}
c_s = 19.6 + const
\end{equation}
The sky background is from \citet{Cuby2000} but refers to a sky area equal to $\pi (0.5 \lambda/D)^2$, where $\lambda$ is the observing wavelength and $D$ is the telescope diameter. The constant term should account for the fact that here we are interested to the noise associated to the sky background, a quantity that depends on the square root of the number of sky photons detected. In practice, we find $const=4.2$~mag.

For what concerns the stellar background, we may use an interpolatory form described in \citet{Langlois2021}, which describes a median observation with SPHERE. Limiting contrast in good conditions might approximately be better by 0.75 mag. If $K_*$ is the star K-band magnitude and $d$ is the star-planet separation in mas, assuming typical values for the number of Detector Integration Time (DIT), $n_{\rm DIT}=48$, for the total exposure time $t_{\rm exp}=72s\times n_{\rm DIT}$, and for the inclination of the system, $\alpha=60^{\circ}$, we can define the following parameters
\begin{align*}
r = & 20.7 - K_*+1.25 \log{t_{\rm exp} /3600}\\
s = & 20.7 +1.25 \log{n_{\rm DIT}/4.6}- K_* +1.25 \log10{t_{\rm exp} /3600}\\
p = & 17.07+ 3.0 (500/d) \log{\alpha/60}+8.75 \log{d/500}\\
q = &-0.5 K_* +18.05+1.25 \log{t_{\rm exp}/3600}+4.375 \log{d/500}\\
c_* = & -2.5 \log{\sqrt{10^{-0.8 p}+10^{-0.8 q}+10^{-0.8 r}+10^{-0.8 s}}}
\end{align*}

For the planet background, we interpolated a formula through the planet-satellite limiting contrast curve derived in \citet{Lazzoni2} for DH Tau. If $x$ is the satellite-planet separation in mas and $\Delta K_p$ is the star planet- contrast in the K-band, then:

\begin{align*}
c_p = & \Delta K_p + (-9.132*10^{-5}*x^2 + 3.85*10^{-2}*x +2.452)
\end{align*}

\begin{figure}
    \centering
    \includegraphics[width=\columnwidth]{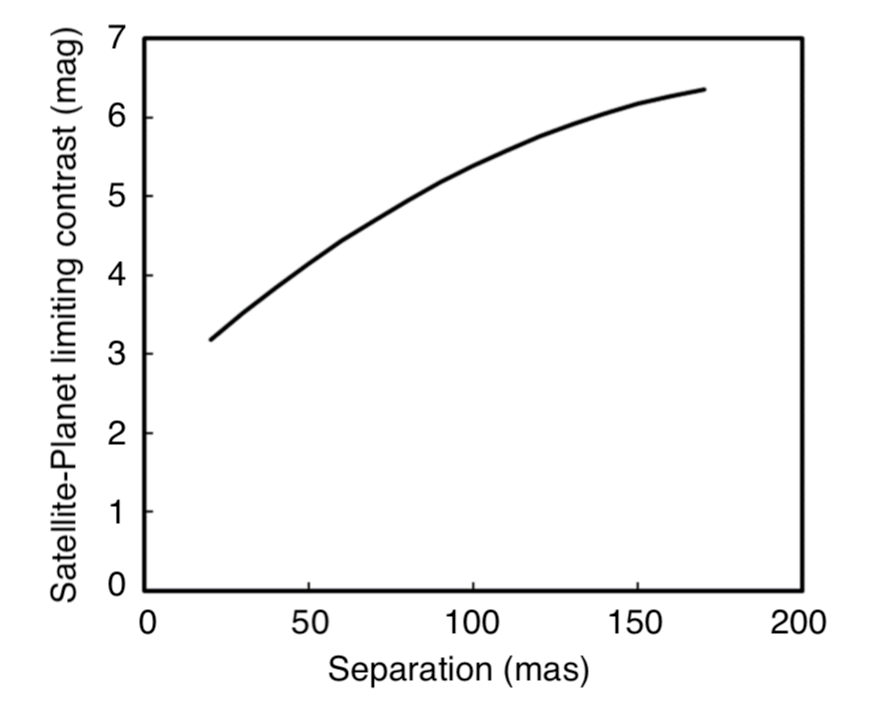}
    \caption{Assumed planet-satellite limiting contrast as a function of separation}
    \label{f:plan_sat_contrast}
\end{figure}

This formula is valid for separation $<220$~mas. At larger separation the contrast due to the planet is negligible with respect to that due to the sky. Figure~\ref{f:plan_sat_contrast} shows the limiting contrast for satellite detection due to the planet background as a function of planet-satellite separation provided by this formula. Note that this sums in quadrature with other sources of background as given by eq. (7), so that the actual limit may be brighter depending on the particular planet we consider. The estimation for the contrast given in this Section are to be considered only as an order of magnitude more than a precise result. For example, for the $\beta$ Pic system, from Figure \ref{betapic_di} we obtain that the minimum detectable mass is $\sim 11.1$ M\textsubscript{Jup}, which correspond to $q=0.87$. From Table \ref{t:prob1}, we retrieved a minimum value for the mass ratio of $0.59$. The discrepancy between the two results is mostly given by the different procedures followed when retrieving the $K_P$ magnitudes for $\beta$ Pic b: from observations in Section 3 and from the mass using interpolated equations in this Section.

\subsubsection{Radial velocities}

Considering the discussion presented in Section 3.2, we assumed that the limiting detectable amplitude of a radial velocity signal by a satellite is $0.1\times 10^{0.2 (K_p-13.5)}$~km/s. This is true for objects with magnitudes close to $K_p\lesssim$ 15 (such as $\beta$ Pic b), whereas for less bright planets we are mostly limited by background noise and the detection probability decreases much more quickly. Thus, the estimates given in Tables \ref{t:prob1} and \ref{t:prob2} are slightly optimistic when considering the dimmest substellar objects. We also assumed that only satellites with periods shorter than 20 yr can be discovered using radial velocities.

We note that when considering the whole sample considered in Table~\ref{t:planet}, the gain using best performing AO but lower transmission of HiRISE with respect to MACAO should be considered object-by-object.

\subsubsection{Transits}

We assumed that transits may be detected if the transit signal corresponds to at least a 20$\%$ dimming of the flux of the planet/BD, as mentioned in Section 3.3. We also assumed that only satellites with periods shorter than 0.1 yr can be discovered using transits.

\subsubsection{Astrometry}

Considering the discussion presented in Section 3.4, we assumed that the limiting detectable amplitude of an astrometric signal by a satellite is 0.3 mas. This corresponds to a $3~\sigma$\ detection within a sequence of observations done with Gravity. We also assumed that only satellites with periods shorter than 20 yr can be discovered using astrometry.

\subsection{Results}

Tables~\ref{t:prob1} and \ref{t:prob2}  give the detection probability and detected satellite parameters for the binary-like and planet-like satellite populations, respectively. As mentioned above, we should multiply these probabilities for the expected frequencies of satellites around planets. 

If we do a survey of these 38 planets and brown dwarfs found by direct imaging, the probability of finding at least one satellite depends on the technique used and on the real satellite population. In Table~\ref{t:det_prob} we give the probability of detecting at least one satellite around one planet/BD by observing the whole sample. The probability of the detection of at least one satellite by observation of the whole sample of 38 planets/BDs, $P_{\rm tot}$, is different from the probability $p_i$ of detecting a satellite around each planet in the sample (reported in Tables ~\ref{t:prob1} and \ref{t:prob2}). The relation between these two probabilities is:
\begin{equation}
P_{\rm tot} = 1 - \prod{1-p_i}
\end{equation}
In all cases, the probabilities do not only depend on the satellite detectability, but also on the actual frequency of satellites in the population considered. We then give two values for $P_{\rm tot}$, for a frequency f=1 (all planets have one satellite belonging to that particular population) and more realistic values obtained considering a lower frequency of f=0.14 for the binary-like population and of f=0.3 for the \textit{planet-like} satellite population. Though the latter two frequencies were chosen rather arbitrarily to give order of magnitude estimates, they were inspired by the the frequency of companions to low-mass stars \citep{Fisher1, Delfosse, Janson3} and the frequency of giant planets around Solar-type stars \citep[see e.g. Table 1 of ][]{Rameau2} In general, the probability of finding at least one satellite is reasonably high if satellites distribute according to the binary distribution, that is, if planets and satellites form like binaries by cloud or disk fragmentation or by capture. It is low, instead, if satellites around planets are distributed as planets around stars, that is if they are formed within a core accretion scenario. Even limiting to the first class of objects, the satellites that can be found around direct imaging planets depends on the technique used (see Table~\ref{t:det_param}). Direct imaging is only sensitive to satellites at rather large separations; astrometry and radial velocities probes regions at intermediate separations; while transits can only reveal satellites very close to the planets. In addition, imaging may only reveal satellites with large mass ratio, and hence only those formed in a {\it binary-like} scenario.

When considering campaigns devoted to searching for satellites, we found that the DI technique is by far the least costly in terms of telescope time. In fact, indirect techniques require tens of observations spread over long time intervals. The observing time required to obtain the temporal series required by RV and transit methods for all the planets in the sample is huge (about 2000-4000 hr on an 8 m telescope). Much less observing time ($\sim 100$~hr) is required by direct imaging, while the request for astrometry is intermediate between the two values, but still considerable (about 1000-2000 hr). On the other hand, in principle a single visit is enough to detect a satellite by DI, though of course a few more are likely required for confirmation and characterization. It can then be useful to consider the probability of detection of at least one satellite for a given observing time (say 100 hr). Here, we only consider the detection of binary-like satellites, because the probability of detecting planet-like satellites within this observing time is always very small. Since for direct imaging we may observe the whole sample,   $P_{\rm tot}$   is still that given in Table~\ref{t:det_prob}, that is  97.2\% if f=1 and 36\% if f=0.14. Within 100 hr we may obtain adequate time series for RV and transits only for a couple of planets; selecting best candidates, using RVs we have $P_{\rm tot}=42$\% if f=1 and 6\% if f=0.14; for transits the probabilities are $P_{\rm tot}=3$\% if f=1 and less than 1\% if f=0.14. For astrometry we might observe four targets, yielding $P_{\rm tot}=64$\% if f=1 and 12\% if f=0.14. This clarifies that DI is the most efficient technique, in terms of observing time, for binary-like planets, with astrometry and radial velocities also being interesting possible alternatives.

From this point of view, notably \citet{Lazzoni2} found a candidate satellite to DH Tau B in their DI survey with a mass ratio $q\sim 0.1$ at $\sim 10$~au from the companion. The parameters for this object is similar to those considered here for the {\it binary-like} population. This is particularly intriguing because this is indeed the target with the highest probability of detection of a satellite through DI among those considered in this paper. While this satellite still needs a final confirmation, it shows the possible expectations for a search of satellites through DI.

We notice that the prospects for a satellite search through DI are even better when using ELT because of the much higher spatial resolution that allows observation of satellites at much closer separations than is possible with 8-m telescopes. As an example, we may estimate that simply because of the higher space resolution, the probability of finding a binary-like satellite around GQ Lup b (if it is there) increases from 17 to 30$\%$ when migrating from VLT to ELT.

\begin{table}
    \centering
    \caption{Probability of detecting at least one satellite and expected number of detected satellites extracted from different distributions with different techniques; f is the frequency of giant satellites per planet}
    \begin{tabular}{lcccccc}
\hline
Technique &	\multicolumn{2}{c}{binary-like} & N$_{\rm det}$ & \multicolumn{2}{c}{planet-like}& N$_{\rm det}$ \\
          & f=1.0 & f=0.14 & f=1.0 & f=1.0 & f=0.3 & f=1.0 \\ 
\hline
Astrometry	& 0.999 & 0.58 & 6.1 & 0.08 & 0.024 & 0.08 \\
Imaging	    & 0.965 & 0.35 & 3.2 & 0.00 & 0.000 & 0.00 \\
RV	        & 0.996 & 0.52 & 5.1 & 0.08 & 0.025 & 0.08 \\
Transits    & 0.433 & 0.08 & 0.6 & 0.06 & 0.019 & 0.06 \\
\hline
\end{tabular}
\label{t:det_prob}
\end{table}

\begin{table}
    \centering
    \caption{Median orbital semimajor axis, $a$, in au and mass ratio $q=m_s/M_P$ expected for detected satellites extracted from different distributions with different techniques}
    \begin{tabular}{lcccccccc}
\hline
Technique & Median(a) & Median(q) & Median(a) & Median(q) \\
          &\multicolumn{2}{c}{binary-like} & \multicolumn{2}{c}{planet-like} \\
\hline
Astrometry & 0.60  & 0.52 & 0.66 & 0.066 \\
Imaging    & 2.68  & 0.62 &      &       \\
RV         & 0.57  & 0.52 & 0.15 & 0.050 \\
Transits   & 0.030 & 0.45 & 0.034& 0.012 \\
\hline
\end{tabular}
\label{t:det_param}
\end{table}

\section{Summary}

In this paper we presented an analysis of the detectability, with different techniques, of satellites around directly imaged planets and brown dwarfs. The term satellite is still vague and could include \textit{planet-like} companions, similar to the moons of the Solar system, or \textit{binary-like} companions, with mass ratio with respect to the planet close to $1$.

Directly imaged exoplanets and brown dwarfs look quite promising for the search of satellites thanks to their mass and wide separation from the host star. These conditions could in principle permit the existence of a zoo of satellites with a wide range of masses and separations from the planet. Moreover, the formation history of DI substellar companions is not yet well established and could be inferred by the properties of any putative satellite.

We showed our complete results for a test system, $\beta$ Pic b, plotting the signals that each satellite, under the simplistic assumptions of circular orbits and most favorable inclination angles, could generate if the planet were observed with different techniques. We then generated two populations of satellites, \textit{planet-like} and \textit{binary-like} respectively, on circular orbits and inclinations distributed according to $sin i$, around a sample of 38 DI exoplanets and brown dwarfs and we analyzed their detectability with different techniques (direct imaging, RV, transits and astrometry). From this preliminary analysis it emerged that \textit{planet-like} satellites around substellar companions have very low chances of being detected at present times with any technique (highest probability values of $\sim 0.01$ for RV and Astrometry). On the other hand, a significant fraction of \textit{binary-like} satellites, if present, would be revealed. For this second category of satellites, given the similar detection probabilities for the four techniques analyzed, direct imaging is the most suitable tool for this kind of survey since it requires less observations, compared to any indirect methods, to confirm a candidate.

We also note that new techniques are being studied for the characterization of exoplanets/BD and the search of satellites, such as spectroastrometry and molecular mapping. Spectroastrometry consists of measuring very accurately the position of the photocenter emitted from an unresolved system companion$+$satellite for each wavelength of a spectrum. This technique could be particularly efficient using instruments such as SPIFFIER$+$ERIS \citep{George,Amico,Kenworthy} and, in the future, JWST/NIRSpec \citep{Birkmann}, SPHERE$+$ \citep{Boccaletti2}, or HARMONI (the IFU of the ELT) \citep{Thatte}. Molecular mapping \citep{Sparks,Konopacky3}, instead, consists in the coupling of AO instruments with high-resolution spectrograph to disentangle the contribution of putative satellites from the central planet/BD. Instruments such as CRIRES$+$ \citep{Follert} and ERIS$+$SPIFFIER \citep{Kenworthy} will be the first to push to the limits of the detection capabilities of this technique, though first analyses were already attempted with SINFONI \citep{Hoeijmakers}.

To conclude, all the techniques mentioned will give outstanding results in the search for satellites thanks to the arrival of new instruments and facilities operating both from ground and space, in the near future.

\section*{Acknowledgements}

This work has been supported by the PRIN-INAF 2019 "Planetary systems at young ages (PLATEA). A.Z. acknowledges support from the FONDECYT Iniciaci\'on en investigaci\'on project number 11190837. We also acknowledge financial support from the ASI-INAF agreement n.2018-16-HH.0. For the purpose of open access, the authors have applied a Creative Commons Attribution (CC BY) licence to any Author Accepted Manuscript version arising from this submission.

\section*{DATA AVAILABILITY}
Data used in this paper is available from the authors upon reasonable
request.




\bibliographystyle{mnras}
\bibliography{bibliography} 




\begin{appendix}
\newpage
\section{Tables}
In the following Tables are shown the detection probabilities and physical properties (semimajor-axis and mass ratio) for satellites around the 38 planets and brown dwarfs considered in the sample. In Table \ref{t:prob1} values are given for the \textit{binary-like} satellite population and in Table \ref{t:prob2} for the \textit{planet-like} population. In each column, from left to right, are listed the name of the planet/BD, the detection technique analyzed, the detection probability, the mean, minimum and maximum semi-major axes, and the mean, minumum and maximum mass ratios for detectable satellites, respectively.
\begin{table*}
    \centering
    \caption{Detection probability and detected satellite parameters for the binary-like satellite population. }
    \begin{tabular}{lcccccccc}
\hline
Planet	&	Method	&	Prob	&	$a_{\rm med}$	&	$a_{\rm min}$	&	$a_{\rm max}$	&	$q_{\rm med}$	&	$q_{\rm min}$	&	$q_{\rm max}$	\\
\hline
1RXS J160929.1-210524 b	&	Astro	&	0.1355	&	0.4776	&	0.0013	&	1.8559	&	0.5580	&	0.0109	&	1.0000	\\
1RXS J160929.1-210524 b	&	RV	&	0.1301	&	0.4684	&	0.0013	&	1.8559	&	0.5588	&	0.0109	&	1.0000	\\
1RXS J160929.1-210524 b	&	Transit	&	0.0119	&	0.0247	&	0.0013	&	0.0537	&	0.4628	&	0.0119	&	1.0000	\\
1RXS J160929.1-210524 b	&	Imaging	&	0.1460	&	8.7624	&	2.7933	&	22.6457	&	0.6975	&	0.4007	&	1.0000	\\
	&		&		&		&		&		&		&		&		\\
2M1207 b	&	Astro	&	0.1406	&	0.4411	&	0.0013	&	1.5860	&	0.5268	&	0.0100	&	1.0000	\\
2M1207 b	&	RV	&	0.1297	&	0.4322	&	0.0013	&	1.5860	&	0.5285	&	0.0100	&	1.0000	\\
2M1207 b	&	Transit	&	0.0100	&	0.0215	&	0.0013	&	0.0462	&	0.4659	&	0.0189	&	0.9999	\\
2M1207 b	&	Imaging	&	0.07745 & 3.2889 & 1.0483 & 8.0126 &  0.7970 & 0.6040 & 1.0000	\\
	&		&		&		&		&		&		&		&		\\
51 Eri b	&	Astro	&	0.0708	&	0.2232	&	0.0013	&	0.4938	&	0.4855	&	0.0183	&	1.0000	\\
51 Eri b	&	RV	&	0.0000	&	0.0017	&	0.0013	&	0.0035	&	0.8701	&	0.6850	&	0.9920	\\
51 Eri b	&	Transit	&	0.0089	&	0.0191	&	0.0013	&	0.0412	&	0.4690	&	0.0260	&	0.9999	\\
51 Eri b	&	Imaging	&	0.0000	&	0.0000	&	0.0000	&	0.0000	&	0.0000	&	0.0000	&	0.0000	\\
	&		&		&		&		&		&		&		&		\\
AB Pic b	&	Astro	&	0.1750	&	0.5908	&	0.0013	&	2.2369	&	0.4952	&	0.0100	&	1.0000	\\
AB Pic b	&	RV	&	0.1709	&	0.5875	&	0.0013	&	2.2369	&	0.4955	&	0.0100	&	1.0000	\\
AB Pic b	&	Transit	&	0.0144	&	0.0294	&	0.0013	&	0.0651	&	0.4522	&	0.0100	&	0.9999	\\
AB Pic b	&	Imaging	&	0.1626	&	5.9725	&	1.0025	&	23.7721	&	0.7507	&	0.4999	&	1.0000	\\
	&		&		&		&		&		&		&		&		\\
beta Pic b	&	Astro	&	0.0915	&	0.2373	&	0.0013	&	0.6502	&	0.4653	&	0.0100	&	1.0000	\\
beta Pic b	&	RV	&	0.0911	&	0.2372	&	0.0013	&	0.6502	&	0.4654	&	0.0100	&	1.0000	\\
beta Pic b	&	Transit	&	0.0135	&	0.0283	&	0.0013	&	0.0617	&	0.4561	&	0.0100	&	0.9999	\\
beta Pic b	&	Imaging	&	0.0102	&	0.5145	&	0.3888	&	0.6502	&	0.7936	&	0.5883	&	1.0000	\\
	&		&		&		&		&		&		&		&		\\
CT Cha b	&	Astro	&	0.1471	&	0.8536	&	0.0585	&	2.2854	&	0.5368	&	0.0317	&	1.0000	\\
CT Cha b	&	RV	&	0.1759	&	0.8144	&	0.0013	&	2.2854	&	0.5302	&	0.0102	&	1.0000	\\
CT Cha b	&	Transit	&	0.0146	&	0.0297	&	0.0013	&	0.0665	&	0.4495	&	0.0100	&	0.9999	\\
CT Cha b	&	Imaging	&	0.2785	&	14.7405	&	3.8388	&	47.3395	&	0.5461	&	0.1330	&	1.0000	\\
	&		&		&		&		&		&		&		&		\\
DH Tau B	&	Astro	&	0.1453	&	0.7422	&	0.0413	&	2.0343	&	0.5242	&	0.0253	&	1.0000	\\
DH Tau B	&	RV	&	0.1641	&	0.7207	&	0.0013	&	2.0343	&	0.5204	&	0.0102	&	1.0000	\\
DH Tau B	&	Transit	&	0.0133	&	0.0268	&	0.0013	&	0.0594	&	0.4549	&	0.0100	&	0.9997	\\
DH Tau B	&	Imaging	&	0.3029	&	13.1787	&	2.7064	&	52.1904	&	0.5705	&	0.1882	&	1.0000	\\
	&		&		&		&		&		&		&		&		\\
eta Tel B	&	Astro	&	0.2210	&	0.9666	&	0.0024	&	3.3369	&	0.4734	&	0.0100	&	1.0000	\\
eta Tel B	&	RV	&	0.2291	&	0.9621	&	0.0013	&	3.3369	&	0.4727	&	0.0100	&	1.0000	\\
eta Tel B	&	Transit	&	0.0188	&	0.0384	&	0.0013	&	0.0971	&	0.4505	&	0.0100	&	0.9999	\\
eta Tel B	&	Imaging	&	0.2693	&	5.6070	&	0.9474	&	19.2398	&	0.5713	&	0.1624	&	1.0000	\\
	&		&		&		&		&		&		&		&		\\
GJ504 b	&	Astro	&	0.2040	&	0.7500	&	0.0055	&	2.6347	&	0.4575	&	0.0100	&	1.0000	\\
GJ504 b	&	RV	&	0.0000	&	0.0000	&	0.0000	&	2.6347	&	0.0000	&	0.0000	&	1.0000	\\
GJ504 b	&	Transit	&	0.0165	&	0.0331	&	0.0013	&	0.0767	&	0.4442	&	0.0100	&	1.0000	\\
GJ504 b	&	Imaging	&	0.0000	&	0.0000	&	0.0000	&	0.0000	&	0.0000	&	0.0000	&	0.0000	\\
	&		&		&		&		&		&		&		&		\\
GQ Lup b	&	Astro	&	0.1770	&	0.9696	&	0.0458	&	2.8819	&	0.5160	&	0.0200	&	1.0000	\\
GQ Lup b	&	RV	&	0.2100	&	0.9343	&	0.0013	&	2.8819	&	0.5086	&	0.0100	&	1.0000	\\
GQ Lup b	&	Transit	&	0.0173	&	0.0351	&	0.0013	&	0.0839	&	0.4498	&	0.0100	&	0.9999	\\
GQ Lup b	&	Imaging	&	0.1474	&	6.5694	&	3.0349	&	12.4799	&	0.5440	&	0.1159	&	1.0000	\\
	&		&		&		&		&		&		&		&		\\
HD1160 c	&	Astro	&	0.2134	&	1.1533	&	0.0379	&	3.7470	&	0.4991	&	0.0128	&	1.0000	\\
HD1160 c	&	RV	&	0.2391	&	1.1272	&	0.0014	&	3.7470	&	0.4952	&	0.0100	&	1.0000	\\
HD1160 c	&	Transit	&	0.0200	&	0.0407	&	0.0013	&	0.1090	&	0.4404	&	0.0100	&	1.0000	\\
HD1160 c	&	Imaging	&	0.1478	&	5.5738	&	2.5189	&	10.8196	&	0.5437	&	0.1330	&	1.0000	\\
	&		&		&		&		&		&		&		&		\\
HD4747 B	&	Astro	&	0.1701	&	0.5899	&	0.0060	&	1.7127	&	0.4471	&	0.0100	&	1.0000	\\
HD4747 B	&	RV	&	0.1732	&	0.5890	&	0.0018	&	1.7127	&	0.4469	&	0.0100	&	1.0000	\\
HD4747 B	&	Transit	&	0.0198	&	0.0411	&	0.0013	&	0.1113	&	0.4413	&	0.0100	&	1.0000	\\
HD4747 B	&	Imaging	&	0.0096	&	0.9684	&	0.3850	&	1.7127	&	0.9459	&	0.8767	&	1.0000	\\
\hline
\end{tabular}
\label{t:prob1}
\end{table*}

\addtocounter{table}{-1}

\begin{table*}
    \centering
    \caption{Cont.}
    \begin{tabular}{lcccccccc}
\hline
Planet	&	Method	&	Prob	&	$a_{\rm med}$	&	$a_{\rm min}$	&	$a_{\rm max}$	&	$q_{\rm med}$	&	$q_{\rm min}$	&	$q_{\rm max}$	\\
\hline
HD19467 B	&	Astro	&	0.2436	&	1.0453	&	0.0015	&	3.8957	&	0.4644	&	0.0100	&	1.0000	\\
HD19467 B	&	RV	&	0.2441	&	1.0451	&	0.0014	&	3.8957	&	0.4644	&	0.0100	&	1.0000	\\
HD19467 B	&	Transit	&	0.0204	&	0.0408	&	0.0013	&	0.1131	&	0.4410	&	0.0100	&	0.9999	\\
HD19467 B	&	Imaging	&	0.0643	&	2.4891	&	0.6407	&	6.5292	&	0.8361	&	0.6773	&	1.0000	\\
	&		&		&		&		&		&		&		&		\\
HD72946 B	&	Astro	&	0.1160	&	0.3747	&	0.0079	&	0.9423	&	0.4616	&	0.0100	&	1.0000	\\
HD72946 B	&	RV	&	0.1243	&	0.3719	&	0.0014	&	0.9423	&	0.4597	&	0.0100	&	1.0000	\\
HD72946 B	&	Transit	&	0.0203	&	0.0408	&	0.0013	&	0.1119	&	0.4476	&	0.0100	&	1.0000	\\
HD72946 B	&	Imaging	&	0.0034	&	0.7204	&	0.5178	&	0.9422	&	0.9476	&	0.8937	&	0.9999	\\
	&		&		&		&		&		&		&		&		\\
HD95086 b	&	Astro	&	0.1295	&	0.5609	&	0.0013	&	1.5295	&	0.5133	&	0.0217	&	1.0000	\\
HD95086 b	&	RV	&	0.0000	&	0.0014	&	0.0013	&	0.0023	&	0.9384	&	0.7992	&	0.9961	\\
HD95086 b	&	Transit	&	0.0097	&	0.0209	&	0.0013	&	0.0443	&	0.4602	&	0.0210	&	0.9998	\\
HD95086 b	&	Imaging	&	0.0000	&	0.0000	&	0.0000	&	0.0000	&	0.0000	&	0.0000	&	0.0000	\\
	&		&		&		&		&		&		&		&		\\
HR2562 B	&	Astro	&	0.1853	&	0.5966	&	0.0013	&	2.0906	&	0.5142	&	0.0100	&	1.0000	\\
HR2562 B	&	RV	&	0.1566	&	0.5729	&	0.0013	&	2.0906	&	0.5236	&	0.0101	&	1.0000	\\
HR2562 B	&	Transit	&	0.0172	&	0.0351	&	0.0013	&	0.0831	&	0.4419	&	0.0100	&	1.0000	\\
HR2562 B	&	Imaging	&	0.0000	&	0.0000	&	0.0000	&	0.0000	&	0.0000	&	0.0000	&	0.0000	\\
	&		&		&		&		&		&		&		&		\\
HR3549 B	&	Astro	&	0.2079	&	0.7957	&	0.0013	&	3.3702	&	0.5110	&	0.0102	&	1.0000	\\
HR3549 B	&	RV	&	0.2015	&	0.7885	&	0.0013	&	3.3702	&	0.5116	&	0.0102	&	1.0000	\\
HR3549 B	&	Transit	&	0.0190	&	0.0383	&	0.0013	&	0.0983	&	0.4423	&	0.0100	&	1.0000	\\
HR3549 B	&	Imaging	&	0.0798	&	4.2501	&	1.9066	&	8.1100	&	0.7312	&	0.4601	&	1.0000	\\
	&		&		&		&		&		&		&		&		\\
HR8799 b	&	Astro	&	0.1489	&	0.4253	&	0.0013	&	1.6633	&	0.5731	&	0.0100	&	1.0000	\\
HR8799 b	&	RV	&	0.0374	&	0.1629	&	0.0013	&	1.3903	&	0.7642	&	0.0584	&	1.0000	\\
HR8799 b	&	Transit	&	0.0102	&	0.0227	&	0.0013	&	0.0486	&	0.4619	&	0.0163	&	0.9998	\\
HR8799 b	&	Imaging	&	0.0000	&	0.0000	&	0.0000	&	0.0000	&	0.0000	&	0.0000	&	0.0000	\\
	&		&		&		&		&		&		&		&		\\
HR8799 c	&	Astro	&	0.1575	&	0.4743	&	0.0013	&	1.8245	&	0.5803	&	0.0100	&	1.0000	\\
HR8799 c	&	RV	&	0.1063	&	0.4161	&	0.0013	&	1.8245	&	0.6182	&	0.0200	&	1.0000	\\
HR8799 c	&	Transit	&	0.0117	&	0.0240	&	0.0013	&	0.0530	&	0.4597	&	0.0125	&	0.9998	\\
HR8799 c	&	Imaging	&	0.0064	&	1.5089	&	0.8261	&	2.4656	&	0.9606	&	0.9208	&	1.0000	\\
	&		&		&		&		&		&		&		&		\\
HR8799 d	&	Astro	&	0.1596	&	0.4937	&	0.0013	&	1.6995	&	0.5327	&	0.0100	&	1.0000	\\
HR8799 d	&	RV	&	0.1317	&	0.4693	&	0.0013	&	1.6995	&	0.5437	&	0.0103	&	1.0000	\\
HR8799 d	&	Transit	&	0.0126	&	0.0261	&	0.0013	&	0.0564	&	0.4589	&	0.0104	&	0.9999	\\
HR8799 d	&	Imaging	&	0.0121	&	1.2168	&	0.8258	&	1.6995	&	0.8792	&	0.7607	&	1.0000	\\
	&		&		&		&		&		&		&		&		\\
HR8799 e	&	Astro	&	0.1124	&	0.3137	&	0.0013	&	0.9663	&	0.5621	&	0.0131	&	0.9999	\\
HR8799 e	&	RV	&	0.0832	&	0.2844	&	0.0013	&	0.9663	&	0.5865	&	0.0199	&	0.9999	\\
HR8799 e	&	Transit	&	0.0117	&	0.0246	&	0.0013	&	0.0530	&	0.4591	&	0.0125	&	0.9999	\\
HR8799 e	&	Imaging	&	0.0007	&	0.8959	&	0.8258	&	0.9663	&	0.9593	&	0.9209	&	1.0000	\\
	&		&		&		&		&		&		&		&		\\
HIP64892 B	&	Astro	&	0.1866	&	0.9721	&	0.0063	&	2.9754	&	0.5065	&	0.0151	&	1.0000	\\
HIP64892 B	&	RV	&	0.1917	&	0.9656	&	0.0013	&	2.9754	&	0.5065	&	0.0140	&	1.0000	\\
HIP64892 B	&	Transit	&	0.0178	&	0.0361	&	0.0013	&	0.0870	&	0.4397	&	0.0100	&	0.9998	\\
HIP64892 B	&	Imaging	&	0.1297	&	6.2910	&	2.5032	&	13.3070	&	0.6611	&	0.3108	&	1.0000	\\
	&		&		&		&		&		&		&		&		\\
HIP65426 b	&	Astro	&	0.1415	&	0.5756	&	0.0013	&	1.8545	&	0.5580	&	0.0225	&	1.0000	\\
HIP65426 b	&	RV	&	0.0163	&	0.0698	&	0.0013	&	0.4690	&	0.7860	&	0.0968	&	1.0000	\\
HIP65426 b	&	Transit	&	0.0120	&	0.0249	&	0.0013	&	0.0540	&	0.4626	&	0.0120	&	1.0000	\\
HIP65426 b	&	Imaging	&	0.0000	&	0.0000	&	0.0000	&	0.0000	&	0.0000	&	0.0000	&	0.0000	\\
	&		&		&		&		&		&		&		&		\\
HIP74865 B	&	Astro	&	0.1834	&	0.9444	&	0.0219	&	2.5711	&	0.4881	&	0.0146	&	1.0000	\\
HIP74865 B	&	RV	&	0.1971	&	0.9305	&	0.0014	&	2.5711	&	0.4869	&	0.0106	&	1.0000	\\
HIP74865 B	&	Transit	&	0.0188	&	0.0380	&	0.0013	&	0.0965	&	0.4445	&	0.0100	&	0.9999	\\
HIP74865 B	&	Imaging	&	0.0027	&	2.5187	&	2.4692	&	2.5711	&	0.6367	&	0.3170	&	0.9999	\\
\hline
\end{tabular}
\end{table*}

\addtocounter{table}{-1}

\begin{table*}
    \centering
    \caption{Cont.}
    \begin{tabular}{lcccccccc}
\hline
Planet	&	Method	&	Prob	&	$a_{\rm med}$	&	$a_{\rm min}$	&	$a_{\rm max}$	&	$q_{\rm med}$	&	$q_{\rm min}$	&	$q_{\rm max}$	\\
\hline
HIP78530 B	&	Astro	&	0.1656	&	0.6080	&	0.0013	&	2.5114	&	0.5767	&	0.0113	&	1.0000	\\
HIP78530 B	&	RV	&	0.1385	&	0.5557	&	0.0013	&	2.5114	&	0.5867	&	0.0113	&	1.0000	\\
HIP78530 B	&	Transit	&	0.0160	&	0.0319	&	0.0013	&	0.0728	&	0.4495	&	0.0100	&	1.0000	\\
HIP78530 B	&	Imaging	&	0.1251	&	11.1579	&	2.7473	&	38.5397	&	0.7941	&	0.5982	&	1.0000	\\
	&		&		&		&		&		&		&		&		\\
HIP79098 B	&	Astro	&	0.1640	&	0.8773	&	0.0176	&	2.5140	&	0.5180	&	0.0222	&	1.0000	\\
HIP79098 B	&	RV	&	0.1871	&	0.8528	&	0.0013	&	2.5140	&	0.5136	&	0.0103	&	1.0000	\\
HIP79098 B	&	Transit	&	0.0157	&	0.0322	&	0.0013	&	0.0732	&	0.4521	&	0.0100	&	1.0000	\\
HIP79098 B	&	Imaging	&	0.1861	&	8.7350	&	2.9283	&	20.4640	&	0.5911	&	0.1748	&	1.0000	\\
	&		&		&		&		&		&		&		&		\\
HIP107412 B	&	Astro	&	0.1286	&	0.3761	&	0.0013	&	1.1828	&	0.5176	&	0.0113	&	1.0000	\\
HIP107412 B	&	RV	&	0.1145	&	0.3651	&	0.0013	&	1.1828	&	0.5225	&	0.0113	&	1.0000	\\
HIP107412 B	&	Transit	&	0.0167	&	0.0340	&	0.0013	&	0.0788	&	0.4482	&	0.0100	&	0.9999	\\
HIP107412 B	&	Imaging	&	0.0000	&	0.0000	&	0.0000	&	0.0000	&	0.0000	&	0.0000	&	0.0000	\\
	&		&		&		&		&		&		&		&		\\
k And b	&	Astro	&	0.1876	&	0.7810	&	0.0056	&	2.5184	&	0.4798	&	0.0100	&	1.0000	\\
k And b	&	RV	&	0.1921	&	0.7775	&	0.0016	&	2.5184	&	0.4796	&	0.0100	&	1.0000	\\
k And b	&	Transit	&	0.0157	&	0.0321	&	0.0013	&	0.0733	&	0.4493	&	0.0100	&	0.9999	\\
k And b	&	Imaging	&	0.1161	&	3.1716	&	1.0000	&	7.0012	&	0.6800	&	0.3500	&	1.0000	\\
	&		&		&		&		&		&		&		&		\\
PDS 70 b	&	Astro	&	0.1247	&	0.5885	&	0.0023	&	1.3972	&	0.5092	&	0.0249	&	1.0000	\\
PDS 70 b	&	RV	&	0.1360	&	0.5760	&	0.0014	&	1.3972	&	0.5072	&	0.0102	&	1.0000	\\
PDS 70 b	&	Transit	&	0.0118	&	0.0249	&	0.0013	&	0.0539	&	0.4615	&	0.0121	&	0.9999	\\
PDS 70 b	&	Imaging	&	0.0000	&	0.0000	&	0.0000	&	0.0000	&	0.0000	&	0.0000	&	0.0000	\\
	&		&		&		&		&		&		&		&		\\
PDS 70 c	&	Astro	&	0.1391	&	0.6691	&	0.0344	&	1.8374	&	0.5217	&	0.0239	&	1.0000	\\
PDS 70 c	&	RV	&	0.1485	&	0.6596	&	0.0015	&	1.8374	&	0.5208	&	0.0158	&	1.0000	\\
PDS 70 c	&	Transit	&	0.0117	&	0.0248	&	0.0013	&	0.0536	&	0.4538	&	0.0122	&	0.9998	\\
PDS 70 c	&	Imaging	&	0.0008	&	2.2825	&	2.2677	&	2.2980	&	0.6878	&	0.3886	&	0.9997	\\
	&		&		&		&		&		&		&		&		\\
PZ Tel B	&	Astro	&	0.2241	&	0.9882	&	0.0144	&	3.4626	&	0.4718	&	0.0100	&	1.0000	\\
PZ Tel B	&	RV	&	0.2328	&	0.9846	&	0.0014	&	3.4626	&	0.4715	&	0.0100	&	1.0000	\\
PZ Tel B	&	Transit	&	0.0194	&	0.0389	&	0.0013	&	0.1008	&	0.4413	&	0.0100	&	0.9999	\\
PZ Tel B	&	Imaging	&	0.1947	&	3.7262	&	0.9426	&	9.5023	&	0.5671	&	0.1500	&	1.0000	\\
	&		&		&		&		&		&		&		&		\\
TYC 7084-794-1 B	&	Astro	&	0.2141	&	0.7346	&	0.0013	&	2.9407	&	0.4858	&	0.0100	&	1.0000	\\
TYC 7084-794-1 B	&	RV	&	0.2005	&	0.7292	&	0.0013	&	2.9407	&	0.4876	&	0.0100	&	1.0000	\\
TYC 7084-794-1 B	&	Transit	&	0.0174	&	0.0358	&	0.0013	&	0.0857	&	0.4465	&	0.0100	&	0.9999	\\
TYC 7084-794-1 B	&	Imaging	&	0.0990	&	2.8320	&	0.4482	&	9.2910	&	0.8078	&	0.6235	&	1.0000	\\
	&		&		&		&		&		&		&		&		\\
TYC 8047-232-1 B	&	Astro	&	0.1640	&	0.5727	&	0.0013	&	2.2259	&	0.5285	&	0.0102	&	1.0000	\\
TYC 8047-232-1 B	&	RV	&	0.1557	&	0.5628	&	0.0013	&	2.2259	&	0.5298	&	0.0102	&	1.0000	\\
TYC 8047-232-1 B	&	Transit	&	0.0142	&	0.0295	&	0.0013	&	0.0648	&	0.4587	&	0.0100	&	1.0000	\\
TYC 8047-232-1 B	&	Imaging	&	0.1445	&	7.6613	&	1.7257	&	24.6086	&	0.7510	&	0.5071	&	1.0000	\\
	&		&		&		&		&		&		&		&		\\
TYC 8998-760-1 b	&	Astro	&	0.1631	&	0.5560	&	0.0013	&	2.2309	&	0.5779	&	0.0108	&	1.0000	\\
TYC 8998-760-1 b	&	RV	&	0.1281	&	0.5035	&	0.0013	&	2.2309	&	0.5947	&	0.0108	&	1.0000	\\
TYC 8998-760-1 b	&	Transit	&	0.0144	&	0.0299	&	0.0013	&	0.0651	&	0.4577	&	0.0100	&	1.0000	\\
TYC 8998-760-1 b	&	Imaging	&	0.0751	&	5.5391	&	1.8922	&	13.5359	&	0.8207	&	0.6451	&	1.0000	\\
	&		&		&		&		&		&		&		&		\\
TYC 8998-760-1 c	&	Astro	&	0.1353	&	0.5960	&	0.0013	&	1.6837	&	0.5244	&	0.0213	&	1.0000	\\
TYC 8998-760-1 c	&	RV	&	0.0028	&	0.0158	&	0.0013	&	0.0734	&	0.8173	&	0.1969	&	0.9998	\\
TYC 8998-760-1 c	&	Transit	&	0.0106	&	0.0229	&	0.0013	&	0.0489	&	0.4611	&	0.0158	&	0.9998	\\
TYC 8998-760-1 c	&	Imaging	&	0.0000	&	0.0000	&	0.0000	&	0.0000	&	0.0000	&	0.0000	&	0.0000	\\
\hline
\end{tabular}
\end{table*}

\addtocounter{table}{-1}

\begin{table*}
    \centering
    \caption{Cont.}
    \begin{tabular}{lcccccccc}
\hline
Planet	&	Method	&	Prob	&	$a_{\rm med}$	&	$a_{\rm min}$	&	$a_{\rm max}$	&	$q_{\rm med}$	&	$q_{\rm min}$	&	$q_{\rm max}$	\\
\hline
TYC 8984-2245-1 b	&	Astro	&	0.1339	&	0.6184	&	0.0013	&	1.7118	&	0.5281	&	0.0243	&	1.0000	\\
TYC 8984-2245-1 b	&	RV	&	0.0023	&	0.0140	&	0.0013	&	0.0600	&	0.8245	&	0.2254	&	1.0000	\\
TYC 8984-2245-1 b	&	Transit	&	0.0108	&	0.0233	&	0.0013	&	0.0499	&	0.4660	&	0.0151	&	1.0000	\\
TYC 8984-2245-1 b	&	Imaging	&	0.0000	&	0.0000	&	0.0000	&	0.0000	&	0.0000	&	0.0000	&	0.0000	\\
	&		&		&		&		&		&		&		&		\\
GSC 6214-210 B	&	Astro	&	0.1599	&	0.7707	&	0.0116	&	2.2332	&	0.5129	&	0.0184	&	1.0000	\\
GSC 6214-210 B	&	RV	&	0.1732	&	0.7571	&	0.0013	&	2.2332	&	0.5116	&	0.0101	&	1.0000	\\
GSC 6214-210 B	&	Transit	&	0.0144	&	0.0291	&	0.0013	&	0.0648	&	0.4501	&	0.0100	&	0.9996	\\
GSC 6214-210 B	&	Imaging	&	0.1979	&	7.8972	&	2.1763	&	20.7695	&	0.6121	&	0.2158	&	1.0000	\\
\hline
\end{tabular}
\end{table*}

\begin{table*}
    \centering
    \caption{Detection probability and detected satellite parameters for the planet-like satellite population}
    \begin{tabular}{lcccccccc}
\hline
Planet	&	Method	&	Prob	&	$a_{\rm med}$	&	$a_{\rm min}$	&	$a_{\rm max}$	&	$q_{\rm med}$	&	$q_{\rm min}$	&	$q_{\rm max}$	\\
\hline
1RXS  J160929.1-210524 b	&	Astro	&	0.0003	&	0.6642	&	0.0766	&	1.4529	&	0.1137	&	0.0318	&	0.8339	\\
1RXS  J160929.1-210524 b	&	RV	&	0.0007	&	0.1329	&	0.0045	&	1.4529	&	0.1042	&	0.0188	&	0.8339	\\
1RXS  J160929.1-210524 b	&	Transit	&	0.0010	&	0.0288	&	0.0045	&	0.0482	&	0.0206	&	0.0119	&	0.4809	\\
1RXS  J160929.1-210524 b	&	Imaging	&	0.0000	&	0.0000	&	0.0000	&	0.0000	&	0.0000	&	0.0000	&	0.0000	\\
	&		&		&		&		&		&		&		&		\\
2M1207  b	&	Astro	&	0.0007	&	0.4628	&	0.0331	&	1.2648	&	0.0719	&	0.0144	&	0.9632	\\
2M1207  b	&	RV	&	0.0014	&	0.1739	&	0.0018	&	1.2648	&	0.0629	&	0.0069	&	0.9632	\\
2M1207  b	&	Transit	&	0.0007	&	0.0227	&	0.0031	&	0.0393	&	0.0310	&	0.0189	&	0.6671	\\
2M1207  b	&	Imaging	&	0.0000	&	0.0000	&	0.0000	&	0.0000	&	0.0000	&	0.0000	&	0.0000	\\
	&		&		&		&		&		&		&		&		\\
51  Eri  b	&	Astro	&	0.0008	&	0.2433	&	0.0285	&	0.4934	&	0.0669	&	0.0188	&	0.7197	\\
51  Eri  b	&	RV	&	0.0000	&	0.0000	&	0.0000	&	0.4934	&	0.0000	&	0.0000	&	0.7197	\\
51  Eri  b	&	Transit	&	0.0006	&	0.0193	&	0.0027	&	0.0345	&	0.0501	&	0.0263	&	0.5353	\\
51  Eri  b	&	Imaging	&	0.0000	&	0.0000	&	0.0000	&	0.0000	&	0.0000	&	0.0000	&	0.0000	\\
	&		&		&		&		&		&		&		&		\\
AB  Pic  b	&	Astro	&	0.0018	&	0.7811	&	0.0515	&	1.7940	&	0.0402	&	0.0086	&	0.9798	\\
AB  Pic  b	&	RV	&	0.0025	&	0.5842	&	0.0017	&	1.7940	&	0.0411	&	0.0031	&	0.9798	\\
AB  Pic  b	&	Transit	&	0.0015	&	0.0326	&	0.0039	&	0.0524	&	0.0117	&	0.0069	&	0.9680	\\
AB  Pic  b	&	Imaging	&	0.0000	&	1.1028	&	1.1028	&	1.1028	&	0.7494	&	0.7494	&	0.7494	\\
	&		&		&		&		&		&		&		&		\\
beta  Pic  b	&	Astro	&	0.0025	&	0.3304	&	0.0053	&	0.6499	&	0.0319	&	0.0068	&	0.8908	\\
beta  Pic  b	&	RV	&	0.0026	&	0.3172	&	0.0027	&	0.6499	&	0.0317	&	0.0038	&	0.8908	\\
beta  Pic  b	&	Transit	&	0.0015	&	0.0310	&	0.0026	&	0.0503	&	0.0149	&	0.0081	&	0.4421	\\
beta  Pic  b	&	Imaging	&	0.0000	&	0.4887	&	0.4887	&	0.4887	&	0.8851	&	0.8851	&	0.8851	\\
	&		&		&		&		&		&		&		&		\\
CT  Cha  b	&	Astro	&	0.0003	&	0.8110	&	0.0876	&	1.8280	&	0.1479	&	0.0373	&	0.9598	\\
CT  Cha  b	&	RV	&	0.0029	&	0.1345	&	0.0025	&	1.8280	&	0.0401	&	0.0039	&	0.9598	\\
CT  Cha  b	&	Transit	&	0.0016	&	0.0361	&	0.0050	&	0.0536	&	0.0101	&	0.0064	&	0.7438	\\
CT  Cha  b	&	Imaging	&	0.0000	&	4.4704	&	4.4704	&	4.4704	&	0.1642	&	0.1642	&	0.1642	\\
	&		&		&		&		&		&		&		&		\\
DH  Tau  B	&	Astro	&	0.0004	&	0.7887	&	0.0719	&	1.6682	&	0.0969	&	0.0286	&	0.9005	\\
DH  Tau  B	&	RV	&	0.0030	&	0.0987	&	0.0014	&	1.6682	&	0.0386	&	0.0023	&	0.9005	\\
DH  Tau  B	&	Transit	&	0.0014	&	0.0302	&	0.0022	&	0.0489	&	0.0142	&	0.0090	&	0.4918	\\
DH  Tau  B	&	Imaging	&	0.0000	&	0.0000	&	0.0000	&	0.0000	&	0.0000	&	0.0000	&	0.0000	\\
	&		&		&		&		&		&		&		&		\\
eta  Tel  B	&	Astro	&	0.0042	&	1.3311	&	0.0161	&	2.8452	&	0.0229	&	0.0058	&	0.9857	\\
eta  Tel  B	&	RV	&	0.0074	&	0.7237	&	0.0069	&	2.8452	&	0.0196	&	0.0020	&	0.9857	\\
eta  Tel  B	&	Transit	&	0.0025	&	0.0484	&	0.0069	&	0.0776	&	0.0037	&	0.0021	&	0.3194	\\
eta  Tel  B	&	Imaging	&	0.0001	&	1.9034	&	1.0727	&	7.3961	&	0.2405	&	0.1518	&	0.9179	\\
	&		&		&		&		&		&		&		&		\\
GJ504  b	&	Astro	&	0.0097	&	1.0044	&	0.0474	&	2.0976	&	0.0108	&	0.0026	&	0.9744	\\
GJ504  b	&	RV	&	0.0000	&	0.0000	&	0.0000	&	2.0976	&	0.0000	&	0.0000	&	0.9744	\\
GJ504  b	&	Transit	&	0.0021	&	0.0395	&	0.0028	&	0.0659	&	0.0074	&	0.0042	&	0.7005	\\
GJ504  b	&	Imaging	&	0.0000	&	0.0000	&	0.0000	&	0.0000	&	0.0000	&	0.0000	&	0.0000	\\
	&		&		&		&		&		&		&		&		\\
GQ  Lup  b	&	Astro	&	0.0007	&	1.1754	&	0.1002	&	2.3452	&	0.0911	&	0.0222	&	0.8262	\\
GQ  Lup  b	&	RV	&	0.0068	&	0.2008	&	0.0037	&	2.3452	&	0.0201	&	0.0016	&	0.8262	\\
GQ  Lup  b	&	Transit	&	0.0023	&	0.0422	&	0.0037	&	0.0707	&	0.0059	&	0.0032	&	0.6655	\\
GQ  Lup  b	&	Imaging	&	0.0001	&	4.5483	&	3.0922	&	7.0493	&	0.1394	&	0.0749	&	0.3595	\\
	&		&		&		&		&		&		&		&		\\
HD1160  c	&	Astro	&	0.0015	&	1.5972	&	0.1798	&	3.0170	&	0.0475	&	0.0132	&	0.9914	\\
HD1160  c	&	RV	&	0.0049	&	0.5412	&	0.0060	&	3.0170	&	0.0292	&	0.0021	&	0.9914	\\
HD1160  c	&	Transit	&	0.0025	&	0.0535	&	0.0060	&	0.0872	&	0.0027	&	0.0015	&	0.2191	\\
HD1160  c	&	Imaging	&	0.0001	&	4.1521	&	2.6365	&	10.1996	&	0.2113	&	0.1137	&	0.6229	\\
	&		&		&		&		&		&		&		&		\\
HD4747  B	&	Astro	&	0.0092	&	0.4226	&	0.0096	&	1.7120	&	0.0145	&	0.0018	&	0.8484	\\
HD4747  B	&	RV	&	0.0070	&	0.2987	&	0.0096	&	1.7118	&	0.0152	&	0.0018	&	0.8484	\\
HD4747  B	&	Transit	&	0.0025	&	0.0538	&	0.0096	&	0.0888	&	0.0027	&	0.0014	&	0.2231	\\
HD4747  B	&	Imaging	&	0.0000	&	0.0000	&	0.0000	&	0.0000	&	0.0000	&	0.0000	&	0.0000	\\
\hline
\end{tabular}
\label{t:prob2}
\end{table*}

\addtocounter{table}{-1}

\begin{table*}
    \centering
    \caption{Cont.}
    \begin{tabular}{lcccccccc}
\hline
Planet	&	Method	&	Prob	&	$a_{\rm med}$	&	$a_{\rm min}$	&	$a_{\rm max}$	&	$q_{\rm med}$	&	$q_{\rm min}$	&	$q_{\rm max}$	\\
\hline
HD19467  B	&	Astro	&	0.0097	&	0.9089	&	0.0054	&	3.5646	&	0.0166	&	0.0020	&	0.9293	\\
HD19467  B	&	RV	&	0.0051	&	0.3409	&	0.0054	&	3.5646	&	0.0217	&	0.0020	&	0.9293	\\
HD19467  B	&	Transit	&	0.0027	&	0.0545	&	0.0054	&	0.0902	&	0.0024	&	0.0013	&	0.1776	\\
HD19467  B	&	Imaging	&	0.0000	&	1.3402	&	0.7194	&	3.1137	&	0.7971	&	0.7304	&	0.9293	\\
	&		&		&		&		&		&		&		&		\\
HD72946  B	&	Astro	&	0.0026	&	0.6255	&	0.0596	&	0.9386	&	0.0246	&	0.0083	&	0.9814	\\
HD72946  B	&	RV	&	0.0061	&	0.4159	&	0.0144	&	0.9386	&	0.0174	&	0.0018	&	0.9814	\\
HD72946  B	&	Transit	&	0.0026	&	0.0554	&	0.0067	&	0.0899	&	0.0025	&	0.0014	&	0.2643	\\
HD72946  B	&	Imaging	&	0.0000	&	0.0000	&	0.0000	&	0.0000	&	0.0000	&	0.0000	&	0.0000	\\
	&		&		&		&		&		&		&		&		\\
HD95086  b	&	Astro	&	0.0004	&	0.4773	&	0.0546	&	1.2067	&	0.1146	&	0.0242	&	0.9472	\\
HD95086  b	&	RV	&	0.0000	&	0.0000	&	0.0000	&	1.2067	&	0.0000	&	0.0000	&	0.9472	\\
HD95086  b	&	Transit	&	0.0007	&	0.0238	&	0.0028	&	0.0363	&	0.0345	&	0.0210	&	0.4676	\\
HD95086  b	&	Imaging	&	0.0000	&	0.0000	&	0.0000	&	0.0000	&	0.0000	&	0.0000	&	0.0000	\\
	&		&		&		&		&		&		&		&		\\
HR2562  B	&	Astro	&	0.0043	&	0.9190	&	0.0064	&	2.0856	&	0.0258	&	0.0053	&	0.9382	\\
HR2562  B	&	RV	&	0.0007	&	0.1957	&	0.0064	&	1.9036	&	0.1112	&	0.0136	&	0.9382	\\
HR2562  B	&	Transit	&	0.0021	&	0.0422	&	0.0044	&	0.0685	&	0.0059	&	0.0034	&	0.3361	\\
HR2562  B	&	Imaging	&	0.0000	&	0.0000	&	0.0000	&	0.0000	&	0.0000	&	0.0000	&	0.0000	\\
	&		&		&		&		&		&		&		&		\\
HR3549  B	&	Astro	&	0.0017	&	0.7073	&	0.0078	&	2.7002	&	0.0593	&	0.0071	&	0.9044	\\
HR3549  B	&	RV	&	0.0009	&	0.2998	&	0.0078	&	2.6890	&	0.0705	&	0.0071	&	0.9044	\\
HR3549  B	&	Transit	&	0.0024	&	0.0489	&	0.0066	&	0.0795	&	0.0037	&	0.0021	&	0.9044	\\
HR3549  B	&	Imaging	&	0.0000	&	2.2563	&	2.2563	&	2.2563	&	0.6777	&	0.6777	&	0.6777	\\
	&		&		&		&		&		&		&		&		\\
HR8799  b	&	Astro	&	0.0012	&	0.4413	&	0.0145	&	1.3470	&	0.0613	&	0.0099	&	0.8706	\\
HR8799  b	&	RV	&	0.0001	&	0.0622	&	0.0145	&	0.8883	&	0.4220	&	0.2438	&	0.8706	\\
HR8799  b	&	Transit	&	0.0008	&	0.0237	&	0.0032	&	0.0416	&	0.0284	&	0.0163	&	0.3968	\\
HR8799  b	&	Imaging	&	0.0000	&	0.0000	&	0.0000	&	0.0000	&	0.0000	&	0.0000	&	0.0000	\\
	&		&		&		&		&		&		&		&		\\
HR8799  c	&	Astro	&	0.0015	&	0.4562	&	0.0050	&	1.4528	&	0.0569	&	0.0093	&	0.9328	\\
HR8799  c	&	RV	&	0.0003	&	0.0646	&	0.0050	&	1.3785	&	0.1704	&	0.0322	&	0.9328	\\
HR8799  c	&	Transit	&	0.0011	&	0.0268	&	0.0016	&	0.0478	&	0.0215	&	0.0125	&	0.7489	\\
HR8799  c	&	Imaging	&	0.0000	&	0.0000	&	0.0000	&	0.0000	&	0.0000	&	0.0000	&	0.0000	\\
	&		&		&		&		&		&		&		&		\\
HR8799  d	&	Astro	&	0.0016	&	0.2950	&	0.0023	&	1.5246	&	0.0652	&	0.0096	&	0.7830	\\
HR8799  d	&	RV	&	0.0008	&	0.0829	&	0.0023	&	1.5193	&	0.0955	&	0.0115	&	0.7830	\\
HR8799  d	&	Transit	&	0.0012	&	0.0292	&	0.0023	&	0.0454	&	0.0175	&	0.0104	&	0.6983	\\
HR8799  d	&	Imaging	&	0.0000	&	0.0000	&	0.0000	&	0.0000	&	0.0000	&	0.0000	&	0.0000	\\
	&		&		&		&		&		&		&		&		\\
HR8799  e	&	Astro	&	0.0011	&	0.2809	&	0.0033	&	0.9655	&	0.0856	&	0.0148	&	0.9761	\\
HR8799  e	&	RV	&	0.0003	&	0.0483	&	0.0033	&	0.6771	&	0.1585	&	0.0261	&	0.9761	\\
HR8799  e	&	Transit	&	0.0011	&	0.0266	&	0.0016	&	0.0430	&	0.0218	&	0.0125	&	0.3804	\\
HR8799  e	&	Imaging	&	0.0000	&	0.0000	&	0.0000	&	0.0000	&	0.0000	&	0.0000	&	0.0000	\\
	&		&		&		&		&		&		&		&		\\
HIP64892  B	&	Astro	&	0.0011	&	1.1176	&	0.1003	&	2.4059	&	0.0675	&	0.0173	&	0.9812	\\
HIP64892  B	&	RV	&	0.0014	&	0.8706	&	0.0057	&	2.4059	&	0.0675	&	0.0099	&	0.9812	\\
HIP64892  B	&	Transit	&	0.0022	&	0.0452	&	0.0039	&	0.0698	&	0.0051	&	0.0030	&	0.4231	\\
HIP64892  B	&	Imaging	&	0.0000	&	3.1587	&	3.1587	&	3.1587	&	0.5464	&	0.5464	&	0.5464	\\
	&		&		&		&		&		&		&		&		\\
HIP65426  b	&	Astro	&	0.0004	&	0.5805	&	0.0091	&	1.4902	&	0.1017	&	0.0233	&	0.9490	\\
HIP65426  b	&	RV	&	0.0000	&	0.0298	&	0.0091	&	0.2000	&	0.8233	&	0.3292	&	0.9490	\\
HIP65426  b	&	Transit	&	0.0012	&	0.0250	&	0.0037	&	0.0456	&	0.0209	&	0.0119	&	0.9490	\\
HIP65426  b	&	Imaging	&	0.0000	&	0.0000	&	0.0000	&	0.0000	&	0.0000	&	0.0000	&	0.0000	\\
	&		&		&		&		&		&		&		&		\\
HIP74865  B	&	Astro	&	0.0011	&	1.3315	&	0.0062	&	2.5680	&	0.0519	&	0.0064	&	0.7037	\\
HIP74865  B	&	RV	&	0.0017	&	0.8260	&	0.0062	&	2.5680	&	0.0464	&	0.0064	&	0.7037	\\
HIP74865  B	&	Transit	&	0.0026	&	0.0483	&	0.0062	&	0.0819	&	0.0039	&	0.0021	&	0.6502	\\
HIP74865  B	&	Imaging	&	0.0000	&	0.0000	&	0.0000	&	0.0000	&	0.0000	&	0.0000	&	0.0000	\\
\hline
\end{tabular}
\end{table*}

\addtocounter{table}{-1}

\begin{table*}
    \centering
    \caption{Cont.}
    \begin{tabular}{lcccccccc}
\hline
Planet	&	Method	&	Prob	&	$a_{\rm med}$	&	$a_{\rm min}$	&	$a_{\rm max}$	&	$q_{\rm med}$	&	$q_{\rm min}$	&	$q_{\rm max}$	\\
\hline
HIP78530  B	&	Astro	&	0.0006	&	0.3904	&	0.0066	&	2.0145	&	0.1301	&	0.0227	&	0.9802	\\
HIP78530  B	&	RV	&	0.0003	&	0.1168	&	0.0066	&	1.6370	&	0.1598	&	0.0237	&	0.9802	\\
HIP78530  B	&	Transit	&	0.0019	&	0.0373	&	0.0022	&	0.0591	&	0.0088	&	0.0049	&	0.9802	\\
HIP78530  B	&	Imaging	&	0.0000	&	0.0000	&	0.0000	&	0.0000	&	0.0000	&	0.0000	&	0.0000	\\
	&		&		&		&		&		&		&		&		\\
HIP79098  B	&	Astro	&	0.0005	&	1.0176	&	0.1077	&	2.0257	&	0.0856	&	0.0228	&	0.7593	\\
HIP79098  B	&	RV	&	0.0031	&	0.1674	&	0.0028	&	2.0257	&	0.0346	&	0.0026	&	0.7593	\\
HIP79098  B	&	Transit	&	0.0019	&	0.0382	&	0.0035	&	0.0606	&	0.0081	&	0.0048	&	0.2713	\\
HIP79098  B	&	Imaging	&	0.0000	&	6.1033	&	6.1033	&	6.1033	&	0.2104	&	0.2104	&	0.2104	\\
	&		&		&		&		&		&		&		&		\\
HIP107412  B	&	Astro	&	0.0020	&	0.4902	&	0.0095	&	1.1827	&	0.0575	&	0.0110	&	0.9275	\\
HIP107412  B	&	RV	&	0.0006	&	0.1436	&	0.0095	&	1.0808	&	0.1183	&	0.0160	&	0.9275	\\
HIP107412  B	&	Transit	&	0.0020	&	0.0401	&	0.0039	&	0.0645	&	0.0071	&	0.0039	&	0.4704	\\
HIP107412  B	&	Imaging	&	0.0000	&	0.0000	&	0.0000	&	0.0000	&	0.0000	&	0.0000	&	0.0000	\\
	&		&		&		&		&		&		&		&		\\
k  And  b	&	Astro	&	0.0023	&	0.9059	&	0.0238	&	2.3888	&	0.0360	&	0.0076	&	0.9282	\\
k  And  b	&	RV	&	0.0051	&	0.3595	&	0.0046	&	2.3888	&	0.0278	&	0.0022	&	0.9282	\\
k  And  b	&	Transit	&	0.0019	&	0.0367	&	0.0034	&	0.0623	&	0.0082	&	0.0048	&	0.7630	\\
k  And  b	&	Imaging	&	0.0000	&	1.9240	&	1.0507	&	4.3007	&	0.6422	&	0.4013	&	0.9282	\\
	&		&		&		&		&		&		&		&		\\
PDS  70  b	&	Astro	&	0.0004	&	0.5038	&	0.0517	&	1.3935	&	0.1124	&	0.0285	&	0.9115	\\
PDS  70  b	&	RV	&	0.0019	&	0.0963	&	0.0023	&	1.3935	&	0.0567	&	0.0060	&	0.9115	\\
PDS  70  b	&	Transit	&	0.0009	&	0.0260	&	0.0023	&	0.0517	&	0.0202	&	0.0121	&	0.7725	\\
PDS  70  b	&	Imaging	&	0.0000	&	0.0000	&	0.0000	&	0.0000	&	0.0000	&	0.0000	&	0.0000	\\
	&		&		&		&		&		&		&		&		\\
PDS  70  c	&	Astro	&	0.0004	&	0.6498	&	0.0662	&	1.4270	&	0.0874	&	0.0273	&	0.7567	\\
PDS  70  c	&	RV	&	0.0020	&	0.0874	&	0.0014	&	1.4270	&	0.0527	&	0.0046	&	0.7567	\\
PDS  70  c	&	Transit	&	0.0010	&	0.0274	&	0.0020	&	0.0441	&	0.0202	&	0.0123	&	0.7547	\\
PDS  70  c	&	Imaging	&	0.0000	&	0.0000	&	0.0000	&	0.0000	&	0.0000	&	0.0000	&	0.0000	\\
	&		&		&		&		&		&		&		&		\\
PZ  Tel  B	&	Astro	&	0.0044	&	1.3655	&	0.0529	&	2.7895	&	0.0212	&	0.0055	&	0.9930	\\
PZ  Tel  B	&	RV	&	0.0076	&	0.8264	&	0.0044	&	2.7895	&	0.0179	&	0.0016	&	0.9930	\\
PZ  Tel  B	&	Transit	&	0.0027	&	0.0489	&	0.0062	&	0.0804	&	0.0033	&	0.0019	&	0.5903	\\
PZ  Tel  B	&	Imaging	&	0.0001	&	2.4945	&	0.9779	&	9.1992	&	0.2024	&	0.1360	&	0.5540	\\
	&		&		&		&		&		&		&		&		\\
TYC  7084-794-1  B	&	Astro	&	0.0090	&	0.8453	&	0.0023	&	2.3515	&	0.0171	&	0.0029	&	0.8821	\\
TYC  7084-794-1  B	&	RV	&	0.0024	&	0.1792	&	0.0023	&	2.1956	&	0.0395	&	0.0035	&	0.8821	\\
TYC  7084-794-1  B	&	Transit	&	0.0020	&	0.0422	&	0.0023	&	0.0685	&	0.0055	&	0.0031	&	0.2619	\\
TYC  7084-794-1  B	&	Imaging	&	0.0000	&	0.8161	&	0.4743	&	1.0038	&	0.8468	&	0.6583	&	0.8719	\\
	&		&		&		&		&		&		&		&		\\
TYC  8047-232-1  B	&	Astro	&	0.0008	&	0.8079	&	0.0467	&	1.7777	&	0.0625	&	0.0162	&	0.9701	\\
TYC  8047-232-1  B	&	RV	&	0.0009	&	0.7175	&	0.0078	&	1.7777	&	0.0626	&	0.0133	&	0.9701	\\
TYC  8047-232-1  B	&	Transit	&	0.0014	&	0.0342	&	0.0034	&	0.0531	&	0.0127	&	0.0070	&	0.8509	\\
TYC  8047-232-1  B	&	Imaging	&	0.0000	&	3.3906	&	2.5373	&	3.3906	&	0.7998	&	0.5458	&	0.7998	\\
	&		&		&		&		&		&		&		&		\\
TYC  8998-760-1  b	&	Astro	&	0.0007	&	0.4509	&	0.0050	&	1.8409	&	0.0972	&	0.0175	&	0.9739	\\
TYC  8998-760-1  b	&	RV	&	0.0003	&	0.0832	&	0.0050	&	0.9590	&	0.1540	&	0.0192	&	0.9739	\\
TYC  8998-760-1  b	&	Transit	&	0.0015	&	0.0324	&	0.0030	&	0.0520	&	0.0114	&	0.0069	&	0.5541	\\
TYC  8998-760-1  b	&	Imaging	&	0.0000	&	0.0000	&	0.0000	&	0.0000	&	0.0000	&	0.0000	&	0.0000	\\
	&		&		&		&		&		&		&		&		\\
TYC  8998-760-1  c	&	Astro	&	0.0004	&	0.4497	&	0.0037	&	1.3363	&	0.1377	&	0.0245	&	0.9764	\\
TYC  8998-760-1  c	&	RV	&	0.0000	&	0.0037	&	0.0037	&	0.0037	&	0.3356	&	0.3356	&	0.3356	\\
TYC  8998-760-1  c	&	Transit	&	0.0009	&	0.0245	&	0.0020	&	0.0434	&	0.0280	&	0.0158	&	0.8493	\\
TYC  8998-760-1  c	&	Imaging	&	0.0000	&	0.0000	&	0.0000	&	0.0000	&	0.0000	&	0.0000	&	0.0000	\\
\hline
\end{tabular}
\end{table*}

\addtocounter{table}{-1}

\begin{table*}
    \centering
    \caption{Cont.}
    \begin{tabular}{lcccccccc}
\hline
Planet	&	Method	&	Prob	&	$a_{\rm med}$	&	$a_{\rm min}$	&	$a_{\rm max}$	&	$q_{\rm med}$	&	$q_{\rm min}$	&	$q_{\rm max}$	\\
\hline
TYC  8984-2245-1  b	&	Astro	&	0.0003	&	0.5682	&	0.0181	&	1.4459	&	0.1142	&	0.0276	&	0.9911	\\
TYC  8984-2245-1  b	&	RV	&	0.0000	&	0.0239	&	0.0181	&	0.0239	&	0.9187	&	0.7906	&	0.9187	\\
TYC  8984-2245-1  b	&	Transit	&	0.0009	&	0.0248	&	0.0030	&	0.0411	&	0.0252	&	0.0150	&	0.9187	\\
TYC  8984-2245-1  b	&	Imaging	&	0.0000	&	0.0000	&	0.0000	&	0.0000	&	0.0000	&	0.0000	&	0.0000	\\
	&		&		&		&		&		&		&		&		\\
GSC  6214-210  B	&	Astro	&	0.0007	&	0.8113	&	0.0904	&	1.7815	&	0.0824	&	0.0187	&	0.9526	\\
GSC  6214-210  B	&	RV	&	0.0028	&	0.1463	&	0.0014	&	1.7815	&	0.0423	&	0.0026	&	0.9526	\\
GSC  6214-210  B	&	Transit	&	0.0015	&	0.0334	&	0.0048	&	0.0532	&	0.0118	&	0.0069	&	0.3818	\\
GSC  6214-210  B	&	Imaging	&	0.0000	&	4.8535	&	4.8535	&	4.8535	&	0.5487	&	0.5487	&	0.5487	\\
\hline
\end{tabular}
\end{table*}
\end{appendix}

\bsp	
\label{lastpage}
\end{document}